\documentclass[fleqn,usenatbib]{mnras}

\usepackage{mathptmx}

\usepackage[T1]{fontenc}

\DeclareRobustCommand{\VAN}[3]{#2}
\let\VANthebibliography\thebibliography
\def\thebibliography{\DeclareRobustCommand{\VAN}[3]{##3}\VANthebibliography}

\usepackage{graphicx}	
\usepackage{amsmath}	
\usepackage{amssymb}	
\usepackage[usenames, dvipsnames]{color}

\newcommand{\rb}[1]{{\color{Black}{{#1}}}}
\newcommand{\rjs}[1]{{\color{Black}{{#1}}}}

\title[Evidence for merger-free co-evolution]{Evidence for non-merger co-evolution of galaxies and their supermassive black holes}

\author[Smethurst, Beckmann et al.]{R. J. Smethurst$^{1}$,\thanks{E-mail: rebecca.smethurst@physics.ox.ac.uk} R. S. Beckmann$^{2}$,\thanks{E-mail: ricarda.beckmann@ast.cam.ac.uk}\thanks{First-authorship is shared between Smethurst \& Beckmann} B.~D.~Simmons$^{3}$, A. Coil$^{4}$, J. Devriendt$^{1}$, Y. Dubois$^{5}$, I. L. Garland$^{3}$, \newauthor C. J. Lintott$^{2}$, G. Martin$^{6,7}$, S. Peirani$^{5,8}$
\\
$^{1}$ Oxford Astrophysics, Department of Physics, University of Oxford, Denys Wilkinson Building, Keble Road, Oxford, OX1 3RH, UK\\ 
$^{2}$ Institute of Astronomy, University of Cambridge, Madingley Road, Cambridge, CB3 0HA, UK \\
$^{3}$ Physics Department, Lancaster University, Lancaster, LA1 4YB, UK \\
$^{4}$ Center for Astrophysics and Space Sciences, University of California, San Diego, 9500 Gilman Dr., MC 0424, La Jolla, CA 92093-0424 \\
$^{5}$ Institut d'Astrophysique de Paris/CNRS, 98 bis blvd Arago 75014 Paris \\
$^{6}$ Korea Astronomy and Space Science Institute, 776 Daedeokdae-ro, Yuseong-gu, Daejeon 34055, Korea\\
$^{7}$ Steward Observatory, University of Arizona, 933 N. Cherry Ave, Tucson, AZ 85719, USA\\
$^{8}$ Universit\'e C\^ote d'Azur, Observatoire de la C\^ote d'Azur, CNRS, Laboratoire Lagrange, Bd de l'Observatoire, CS 34229, 06304 Nice Cedex 4, France \\
}

\date{Accepted 2023 February 13. Received 2023 February 2; in original form 2022 November 24}

\pubyear{2022}

\begin{document}
\label{firstpage}
\pagerange{\pageref{firstpage}--\pageref{lastpage}}
\maketitle

\begin{abstract}
Recent observational and theoretical studies have suggested that supermassive black holes (SMBHs) grow mostly through non-merger (`secular') processes. Since galaxy mergers lead to dynamical bulge growth, the only way to observationally isolate non-merger growth is to study galaxies with low bulge-to-total mass ratio (e.g.  $B/T<10\%$). However, bulge growth can also occur due to secular processes, such as disk instabilities, making disk-dominated selections a somewhat incomplete way to select merger-free systems. Here we use the Horizon-AGN simulation to select simulated galaxies which have not undergone a merger since $z=2$, regardless of bulge mass, and investigate their location on typical black hole-galaxy scaling relations in comparison to galaxies with merger dominated histories. While the existence of these correlations has long been interpreted as co-evolution of galaxies and their SMBHs driven by galaxy mergers, we show here that they persist even in the absence of mergers. We find that the correlations between SMBH mass and both total mass and stellar velocity dispersion are independent of B/T ratio for both merger-free and merger-dominated galaxies. In addition, the bulge mass and SMBH mass correlation is still apparent for merger-free galaxies, the intercept for which is dependent on B/T. Galaxy mergers reduce the scatter around the scaling relations, with merger-free systems showing broader scatter. We show that for merger-free galaxies, the co-evolution is dominated by radio-mode feedback, and suggest that the long periods of time between galaxy mergers make an important contribution to the co-evolution between galaxies and SMBHs in all galaxies.
\end{abstract}




\section{Introduction}

The strong correlations that are found between supermassive black hole (SMBH) mass \& velocity dispersion \citep{magorrian98, merritt01, hu08, kormendy11a, mcconnell13, vandenbosch16, batiste17, baldassare20}, SMBH mass \& bulge stellar mass \citep{marconi03, haringrix04, saglia16, sahu19, zhao21} and SMBH \& total stellar mass \citep{cisternas11, simmons13, reines15, davis19, sahu19, ding20, bennert21} suggest that galaxies co-evolve with their central SMBHs \citep[][]{silk98, granato04}. Since galaxy mergers can grow both dispersion supported bulges and SMBHs through redistribution of angular momentum, these correlations have long been interpreted as evolution for co-evolution driven by a few galaxy mergers within a Hubble time \citep{peng07, hopkins08b, jahnke11, heckmanbest14}. Cosmological simulations are also able to reproduce these observed correlations between host galaxy properties and their SMBHs, using a variety of physical models \citep[see][for a recent comparison]{Habouzit2021}.

However, a flurry of new results, both observational and theoretical, have suggested that galaxy mergers may not be the dominant mechanism powering this co-evolution.  An internal, secular co-evolution of galaxies and their SMBHs has been suggested, particularly in works studying lower mass galaxies \citep[e.g][]{greene10b, jiang11b, cisternas11, simmons11,kocevski12,greene20,baldassare20}. In an attempt to isolate the merger-free evolutionary pathway \citet*{ssl17} investigated 101 active galactic nuclei (AGN) hosted by strongly disk-dominated\footnote{Hereafter we will use the term ``bulgeless'' to refer to a strongly disk-dominated system either lacking a classic bulge, or having a bulge-to-total ratio, B/T $< 0.1$.} galaxies which are assumed to be merger free. This assumption is motivated by the fact that simulations have consistently shown that mergers with mass ratios larger than 10:1 will form a classical, pressure supported bulge \citep{walker96, hopkins12, tonini16}. Additionally, \citet{martig12} showed that galaxies which are clearly disk dominated (bulge-to-total ratio, $B/T<0.1$) have likely had a calm accretion history, evolving in the absence of major or minor mergers since $z\sim2$. 

\citet*{ssl17} found that the SMBHs in their bulgeless galaxies were 2 orders of magnitude more massive than predicted by the bulge mass-SMBH mass scaling relation. However, \citet*{ssl17} also found their merger-free sample lay on the typical scaling relation between SMBH and total stellar mass of the galaxy, suggesting that major galaxy mergers do not play a significant role in controlling the co-evolution of galaxies and SMBHs. This work was followed up by \cite{martin18} with the Horizon-AGN simulation, who also found that simulated galaxies with $B/T<0.1$ were significantly offset from the typical bulge-SMBH mass relation, and in addition that only $35\%$ of the cumulative growth of SMBHs over the past $\sim$ 12 billion years (since $z \sim 3$) could be attributed to mergers. Similarly in the EAGLE simulation \cite{mcalpine20} found that while galaxy mergers increase the luminosity of AGN, this does not lead to substantial cumulative SMBH growth, finding that by $z=0$ on average no more than $15\%$ of SMBH mass comes from the enhanced accretion rates triggered via a merger. This confirmed results from \cite{gabor15}, who reported a luminosity spike for AGN in merging galaxies, but contradicts results from \cite{bellovary13}, who reported that mergers do not substantially enhance fuelling of the central AGN. The exact impact of galaxy mergers on AGN activity, and its potential impact on the host galaxy, is therefore still under discussion, but a consensus has emerged that SMBH mass growth is not dominated by galaxy mergers.

Simulations have also shown that the majority of bulge growth does not occur due to mergers. For example, \cite{parry09} find in the Millenium simulation that $<\sim 20\%$ of the stellar mass in bulges is built by mergers (in galaxies with total stellar masses $ <5 \times 10^{10} \ \rm{M}_{\odot}$), with the majority instead built through disk instabilities. \cite{martig12} also find that galaxies with the highest bulge Sérsic index tend to have histories of intense gas accretion and disk instabilities rather than active mergers. Similarly \cite{gargiulo17} find in their SAG simulations that $87\%$ of stars in bulges of Milky Way-like galaxies are present due to disc instability events, rather than mergers. More recently, \cite{du21} find that the evolution of bulge-dominated galaxies is not dominated by mergers using the TNG50 simulation (unlike kinematic "slow rotator" elliptical galaxies which are created by mergers; \citealt{martin18b}). These simulation results are supported by the observational study of \cite{bell17} who found that two of their three galaxies with massive `classical' bulges have stellar halos which are inconsistent with a merger origin, suggesting their bulges have been built through secular processes.

Given the growing amount of evidence that mergers may not be as dominant as first thought in driving either SMBH or bulge growth, this raises the question of what else could cause the correlations between galaxy properties and SMBH mass discussed above. Do galaxy merger-free processes also lead to co-evolution of galaxies and their SMBHs? In this study we therefore make use of the Horizon-AGN simulation suite\footnote{https://www.horizon-simulation.org/} to test whether galaxy-SMBH co-evolution is occurring across a large simulated galaxy population due to non-merger processes by investigating the classic scaling relations of $M_{BH}-M_{\rm{bulge}}$, $M_{BH}-M_{\rm{*}}$ and $M_{BH}-\sigma_{*}$. \rb{Horizon-AGN is a large-scale galaxy evolution simulation which tracks galaxy evolution from cosmic dawn to redshift $z=0$. Horizon-AGN is a tried and tested simulation able to replicate a wide range of observations from across the galaxy and SMBH populations, including the galaxy mass function and  cosmic star formation history distribution \cite[e.g.][]{Kaviraj2017}, and SMBH correlations and luminosity functions \citep[for example][]{volonteri16,Habouzit2022}.}

We describe the Horizon-AGN simulation in Section \ref{sec:HAGN}, the selection of galaxies from the simulation in Section~\ref{sec:galcat}, the selection of merger-free and merger-dominated samples in Section~\ref{sec:galsample}, and the calculation of stellar velocity dispersions in Section~\ref{sec:veldisp}. We show and discuss our results in Section \ref{sec:results}, and we conclude in Section \ref{sec:conclusions}.

\section{Simulation Data}\label{sec:sim}

\subsection{Horizon-AGN}\label{sec:HAGN}

Horizon-AGN is a hydrodynamical simulation of a $100 \rm \ Mpc^3 \ h^{-1}$ cosmological volume run to redshift $z=0$. It was presented in detail in \cite{dubois14} so here we only briefly reiterate key features.

Horizon-AGN was produced using the adaptive mesh refinement code \textsc{ramses} \citep{teyssier02} with a  WMAP-7 $\Lambda$CDM cosmology \citep{komatsu11}: total matter density $\Omega_m=0.272$, dark energy density $\Sigma_\Lambda = 0.728$, amplitude of the matter power spectrum $\sigma_8 = 0.81$, baryon density $\Omega_b = 0.045$, Hubble constant $H_0 = 70.4 \rm \ km s^{-1} \ Mpc^{-1}$ and spectral index $\rm n_s = 0.967$. The size of the simulation box is $100 \sim\rm{Mpc}\rm{h}^{-1}$ (comoving), refined on a root grid of $1024^3$, then adaptively refined up to a maximum resolution of $\Delta x = 1$ proper kpc using a quasi-Lagrangian refinement criterion: cells are (de)refined when the mass in the cell is above (below) 8 times the mass resolution of $M_{\rm DM} = 8.27 \times 10^7 \rm \ M_\odot$ for dark matter (DM) and $~ 2 \times 10^6 \rm \ M_\odot$ for stars. Horizon-AGN includes prescriptions for gas cooling including the contribution from metals released by supernova feedback, star formation and stellar feedback, background UV heating as well as black hole formation, accretion and feedback. Star formation is modelled according to a Schmidt law \cite{kennicutt98} with an efficiency of 0.01, using a Salpeter initial mass function \citep{salpeter55}. Stellar feedback is modelled to include stellar winds, type Ia and type II supernovae \citep{dubois08,Kimm15}. 

BH are formed with an initial seed mass of $10^5 \rm \ M_\odot$ in cells that exceeds the density threshold for star formation ($n_0 = 0.1 \rm \ H cm^{-3}$). BH seed formation is stopped at $z=1.5$. To avoid multiple BH forming in the same galaxies, BH formation is not permitted within a 50 comoving kpc exclusion zone around existing BH. 

BH accretion and feedback are modelled as in \citet{dubois12}. BH gas accretion is modelled via the Bondi-Hoyle-Lyttleton formalism $\dot{M}_{\rm BH} = 4 \pi \alpha G^2 M_{\rm BH}^2 \bar{\rho} / (\bar{\rm c}_{\rm s}^2 + \bar{u}^2)^{3/2} $, capped at the Eddington accretion rate $\dot{M}_{\rm Edd}$, where $\alpha$ is a dimensionless boost factor, $\rm M_{\rm BH}$ is the BH mass, G is the gravitational constant, and $\bar{\rho}$, $\bar{c}_s$ and $\bar{u}$ are the average gas density, sound speed and gas velocity. Following \citet{booth10}, $\alpha=  (\rho / \rho_0  )^2$ if $\rho > \rho_0$ and $\alpha=1$ otherwise. AGN feedback energy is released as $\dot{E}_{\rm AGN} = \epsilon_f \epsilon_r \dot{M}_{\rm BH} c^2$ where $\epsilon_r=0.1$ is the radiative efficiency, $c$ is the speed of light and $\epsilon_f$ is an efficiency factor. At high accretion rates ($\chi = \dot{M}_{BH} / \dot{M}_{\rm Edd} > 0.01$, `quasar mode'), energy is injected as isotropic thermal energy, using $\epsilon_f=0.15$. At low accretion rates ($\chi \leq 0.01$, `jet mode') energy is released in bi-conical outflows with $\epsilon_f = 1$. BHs are not pinned to the centres of galaxies but allowed to move freely with a drag force \citep{ostriker99} applied to account for the unrealistic motions and spurious oscillations which arise from unresolved dynamical friction forces. BHs merge when located within 4 kpc of each other, and when their relative velocity is smaller than the escape velocity of the binary.

\subsection{Galaxy catalogue}\label{sec:galcat}

To identify galaxies and dark matter halos in Horizon-AGN, we use \textsc{adaptahop} \citep{aubert04,tweed2009} with 20 neighbours, a local density threshold of $\rho_{\rm t} = 178 $ times the average dark matter density and a force softening of 2 kpc. A minimum particle number cut of 50 DM or star particles is enforced, leading to a minimum galaxy stellar mass of  $M_{*}\approx 10^{8.5} \rm M_\odot$, where $M_{*}$ is the total mass of all star particles associated with a given galaxy as identified by \textsc{adaptahop}. Bulge masses are taken from \citet{volonteri16}, and were computed using two Sersic profiles: one with $n=1$ for the disc component, and one with the best fit of  $n = 1,2,3$ or 4 for the bulge component.

As BH are free to move within the simulation volume, they are not automatically identified with a given host galaxy. To associate BH with galaxies, we use a set of two spatial criteria: a BH is assigned to a galaxy if it is located within 10\% of the galaxy's DM host halo virial radius, and also located within two effective radii of the galaxy itself. If more than one BH meets both criteria, the most massive object is retained as the central BH \citep[see][for details]{volonteri16}. Galaxy mergers are identified using galaxy merger trees constructed from the galaxy catalogues for $z<6$. Snapshots are on average spaced  every 130 Myr. We use the merger trees to identify major (mass ratios $> 1 : 4$) and minor (mass ratios $1 : 4$ to $1 : 10$) galaxy mergers for each galaxy \citep[see][for details]{martin18}. BH mergers are identified on-the-fly during the simulation. During each BH-BH merger, the less massive BH is considered to merge into the more massive one. 

\begin{figure*}
    \begin{center}
	\includegraphics[width=0.49\textwidth]{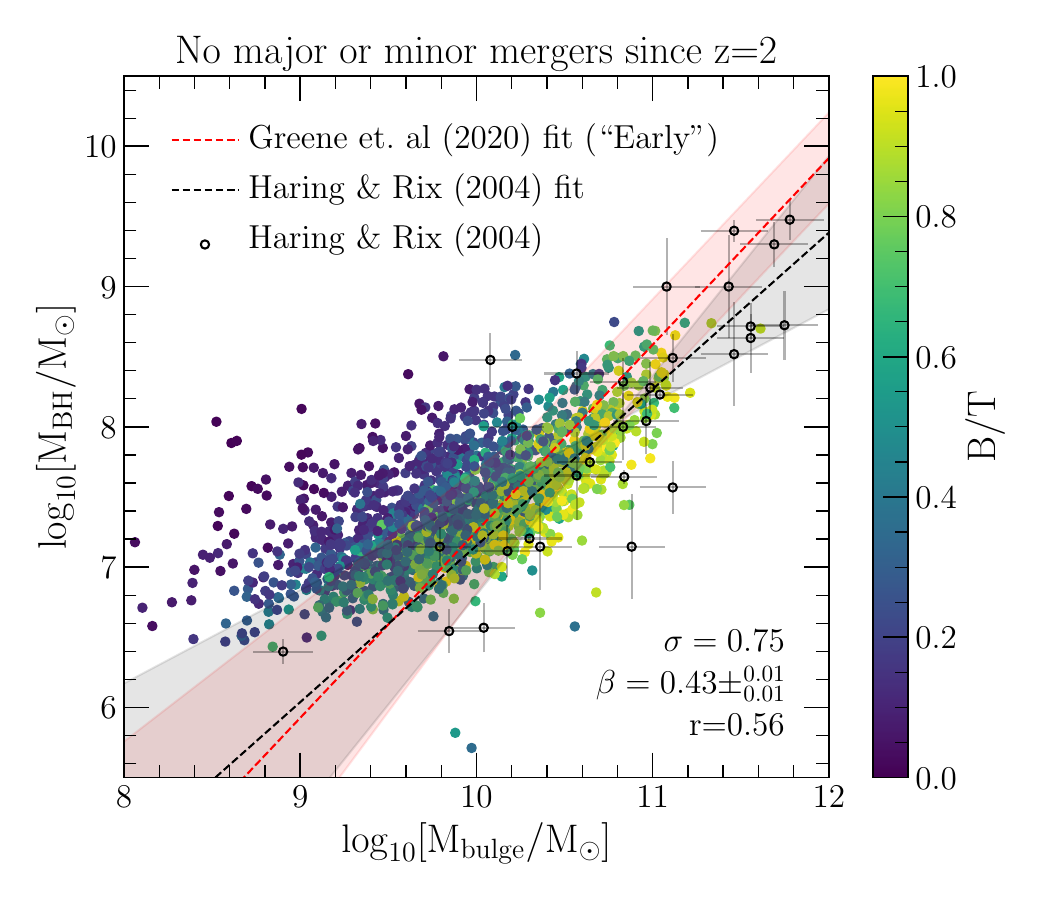}
	\includegraphics[width=0.49\textwidth]{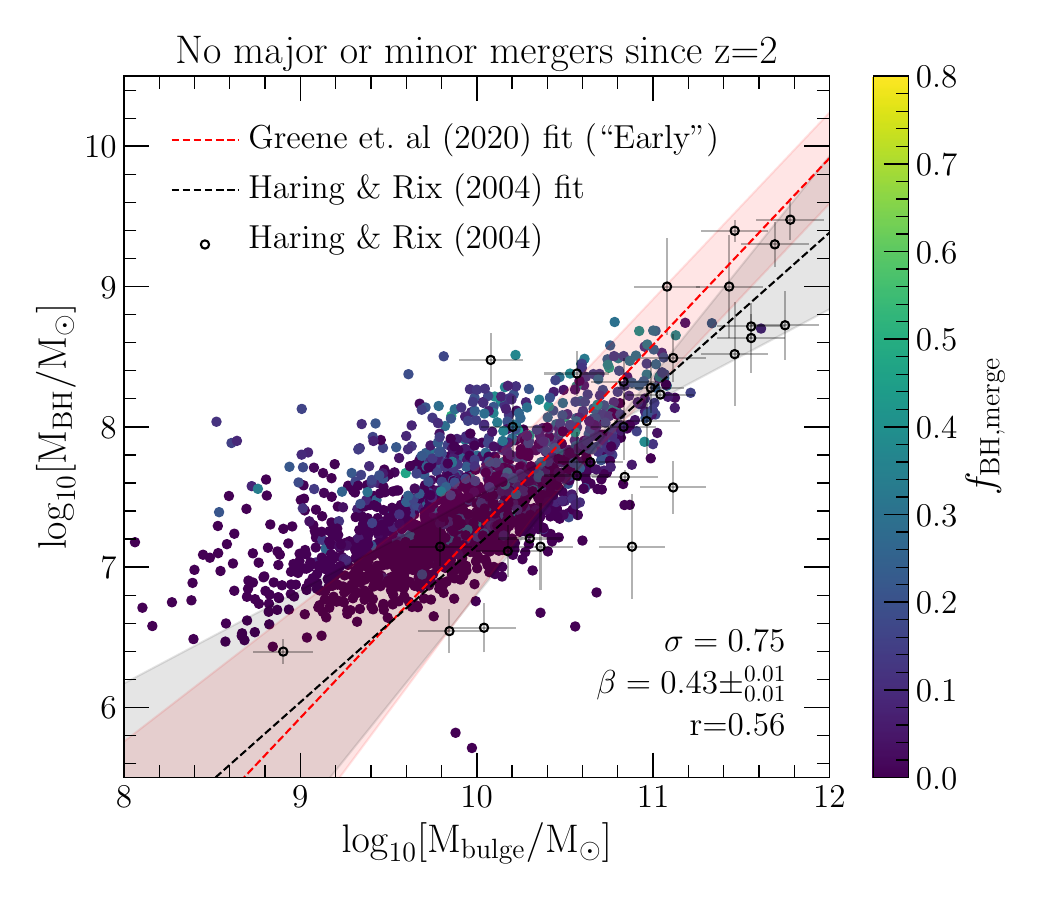}
	\includegraphics[width=0.49\textwidth]{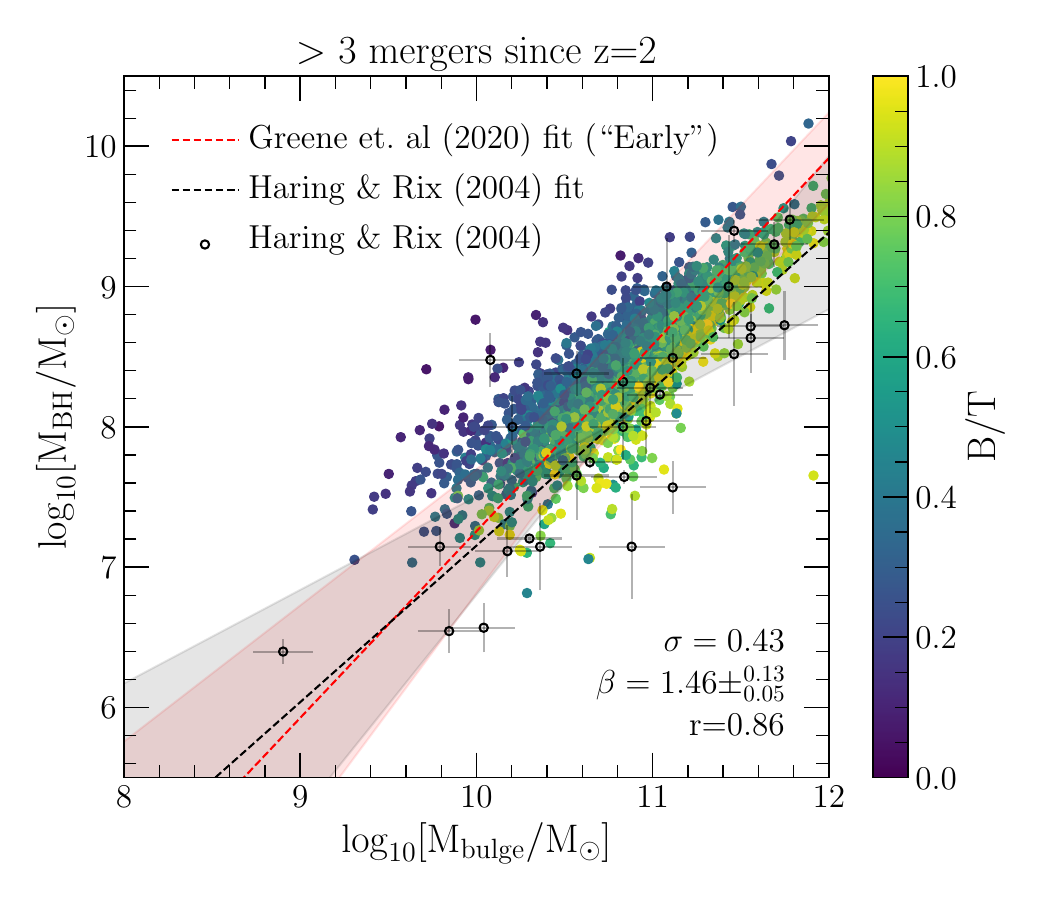}
	\includegraphics[width=0.49\textwidth]{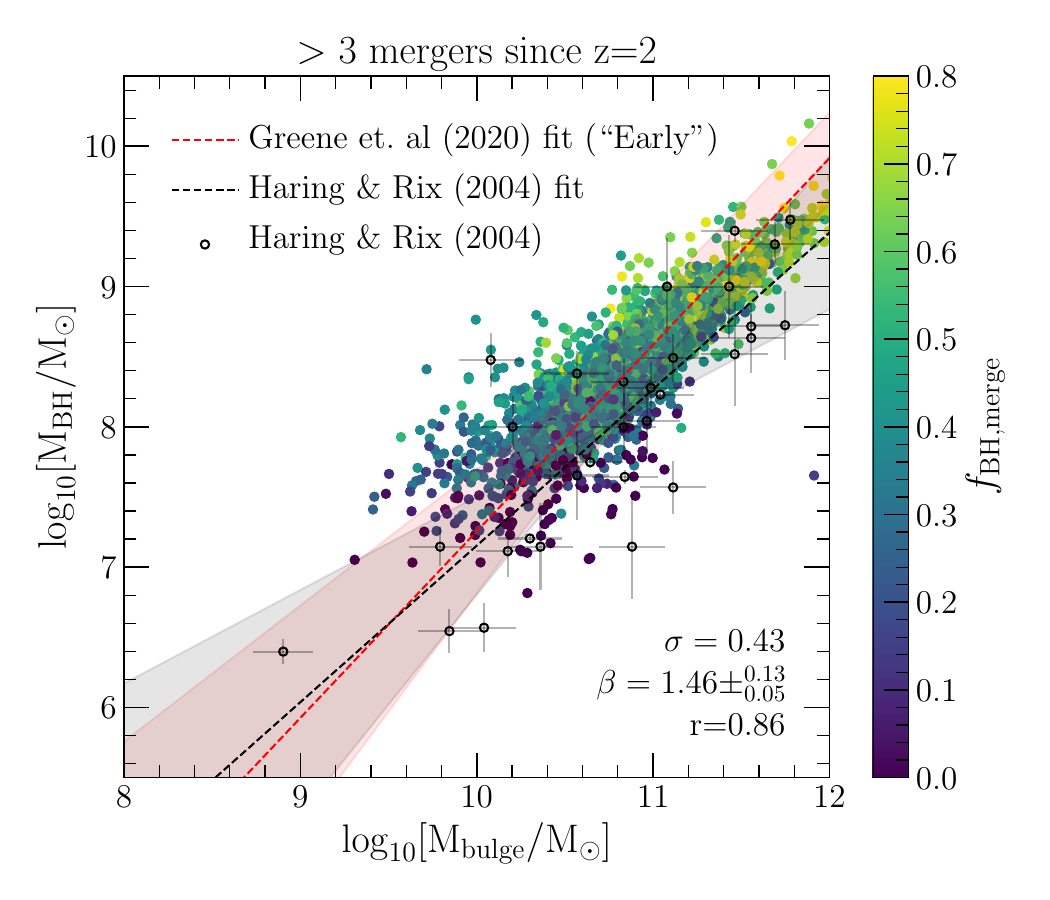}
	\vspace{-1em}
    \caption{Bulge stellar mass against SMBH mass for different subsets of the Horizon-AGN sample. In each panel the black points are those early-type galaxies from \protect\citet[][assumed $B/T=1$]{haringrix04} with the fit shown by the black dashed line (the shaded region shows $\pm1\sigma$). In addition we show the fit to early-type galaxies from \protect\cite{greene20} with the red dashed line (once again the shaded region shows $\pm1\sigma$). The top panels show galaxies which have had neither a major or minor merger since $z=2$ coloured by their bulge-to-total mass ratio ($B/T$; left) and by the fraction of the SMBH mass which was built by BH mergers ($f_{\rm{BH,merge}}$; right). The bottom panels show galaxies which have had more than 3 major or minor mergers since $z=2$ and are coloured by bulge-to-total ratio (left) and by the fraction of the SMBH mass which was built by BH mergers ($f_{\rm{BH,merge}}$; right). The same colour scales have been used on corresponding colour bars across each panel for ease of comparison. In the bottom right corner of each panel we provide the Pearson correlation coefficient, given by the $r$ value, along with the slope of the fit to the simulation data given by the $\beta$ value, and the standard deviation of the points around that fitted slope by the $sigma$ value. A value of $r=+1$ indicates a strong positive linear correlation, whereas a value of $r=0$ indicates no correlation. A larger value of $\beta$ indicates a steeper correlation, and a larger value of $\sigma$ indicates higher scatter around the correlation. A higher scatter is seen for the \textsc{merger-free} sample, but the correlation between bulge mass and SMBH mass is still present ($r=0.78$) with the intercept set by the $B/T$ ratio.}
    \label{fig:bulgefig}
    \end{center}
\end{figure*}

\subsection{Galaxy sample selection}\label{sec:galsample}
Central BHs were identified and bulge masses computed for a total galaxy sample of 6892 galaxies at $z=0.0556$ (the average redshift of the observed `bulgeless' galaxy sample of \citealt{ssl17} for ease of comparison). Out of this sample, we selected two subsamples based on the total (both minor, 10:1 and above, and major, 3:1 and above) number of galaxy mergers experienced by the galaxy since $z=2$. 1801 galaxies ($26\%$) were identified as not having had a merger since $z=2$. We will refer to these galaxies as the \textsc{merger-free} sample. As a comparison sample, we also identified $1271$ galaxies ($18\%$) which have undergone more than 3 mergers (either major or minor) since $z=2$. We will refer to these galaxies as the \textsc{merger-dominated} sample. The redshift cut off at $z=2$ is motivated by both the need for hierarchical structure formation in the early Universe, as per $\Lambda$CDM, and the studies of \citet{martig12} and \citet{martin18} who showed that galaxies with bulge-to-total ratios of $< 0.1$ at $z\sim0$ have not had a merger since at least $z\sim2$. This redshift cutoff also coincides with the peak of star formation density \citep{madau98}. \rjs{The other 3855 galaxies ($56\%$; which we do not investigate here) have a mean of 1.5 mergers since $z=2$ ($0.7$ major mergers on average and $0.8$ minor mergers on average). They are therefore likely evolving through a mix of galaxy mergers and non-merger processes.}

Throughout this study we will compare our samples to the observed galaxy sample fits of \cite{haringrix04}, \cite{mcconnell13} and \cite{greene20}. This facilitates comparison to previous studies, showcases the uncertainty in observed relations, and offers multiple comparisons to different galaxy populations. For example, the canonical relation of \citeauthor{haringrix04} focuses on black hole-galaxy stellar mass relations in early-type galaxies, whereas \citeauthor{greene20} offers an updated fit to both early- and late-type galaxies with an emphasis on low-mass systems (unlike \citeauthor{mcconnell13}, whose $M_{BH}-\sigma_\ast$ relation has no deliberately biased mass selection). 

\begin{figure*}
    \begin{center}
	\includegraphics[width=0.49\textwidth]{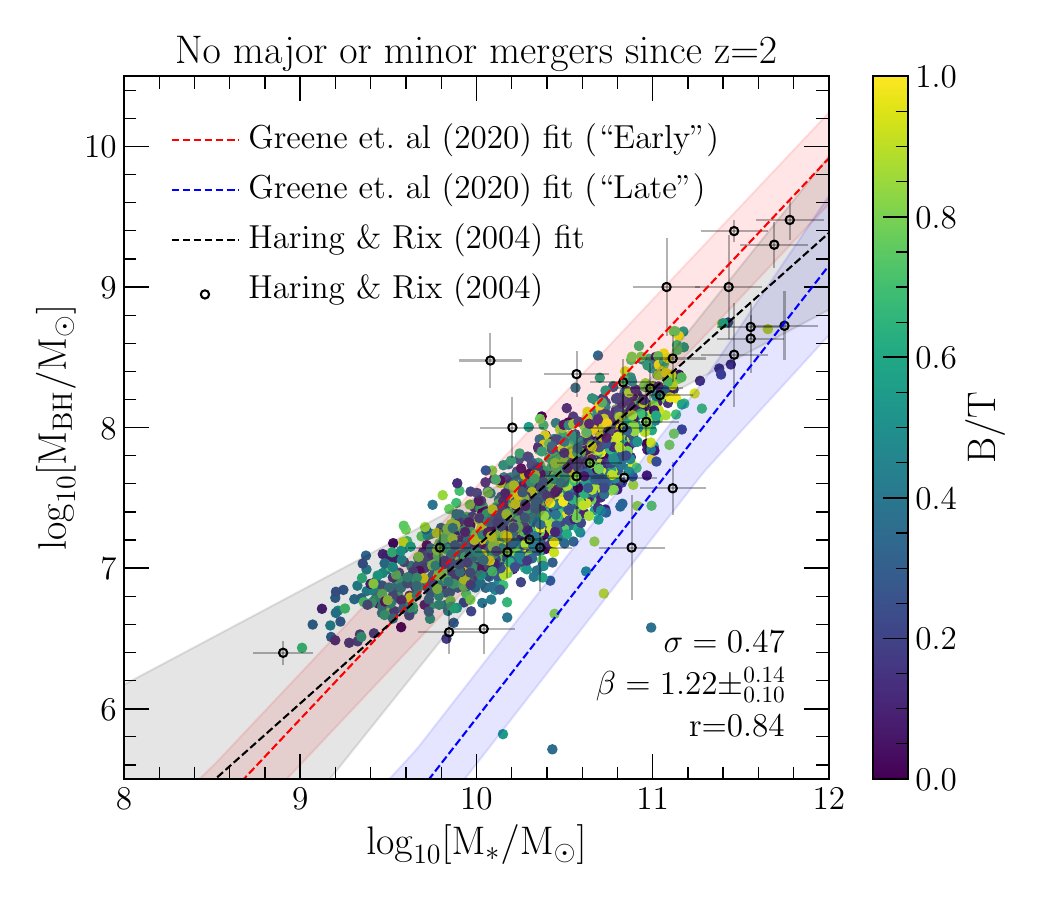}
	\includegraphics[width=0.49\textwidth]{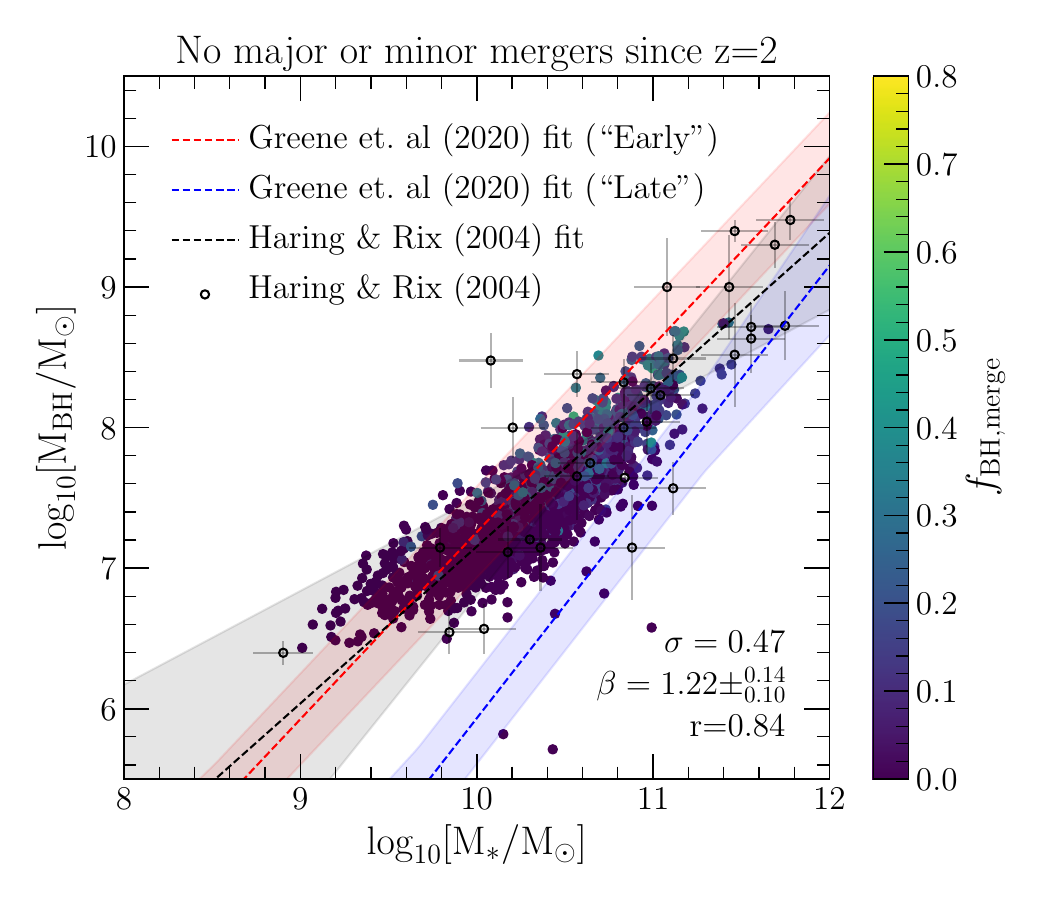}
	\includegraphics[width=0.49\textwidth]{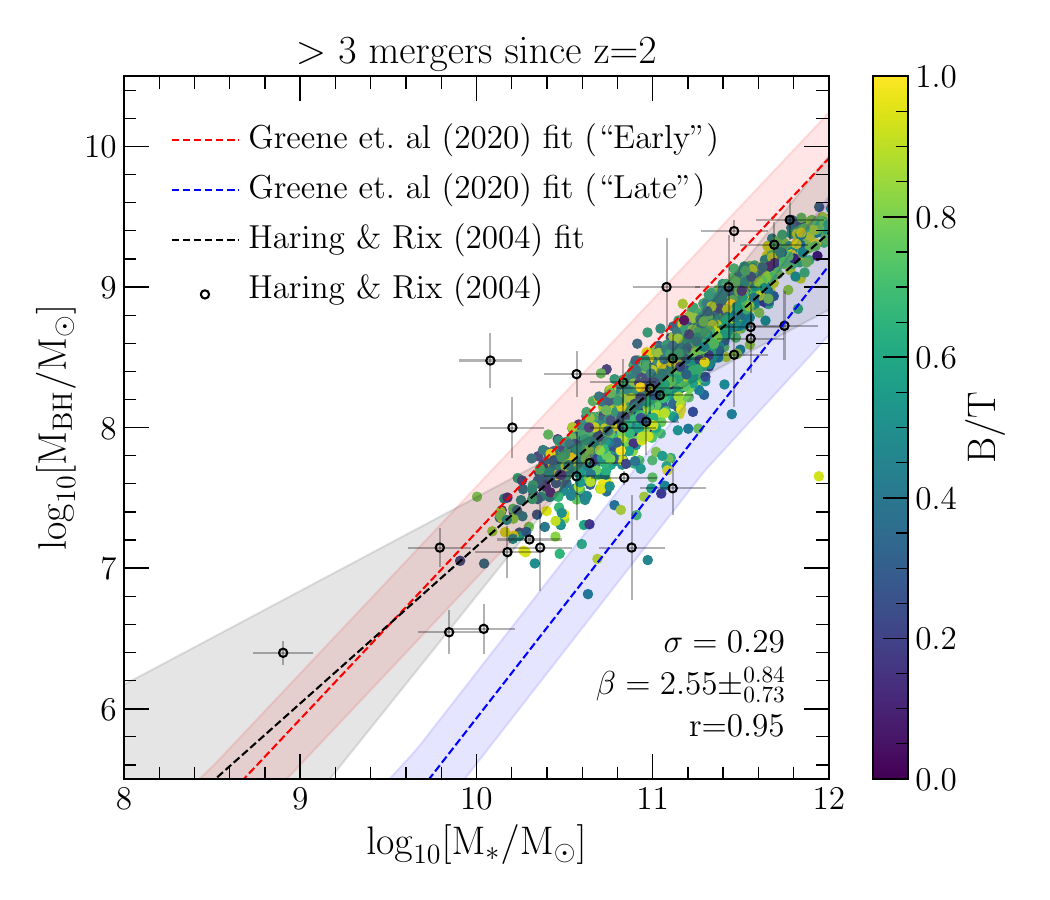}
	\includegraphics[width=0.49\textwidth]{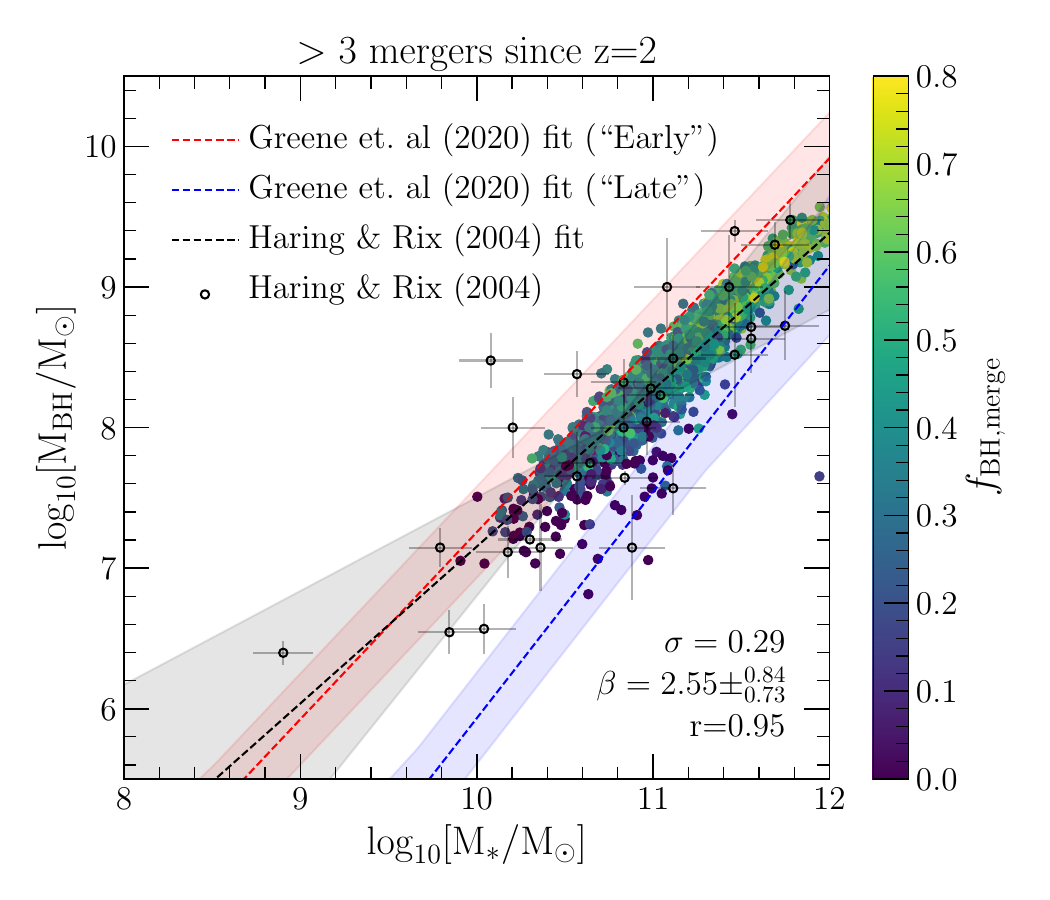}
	\vspace{-1em}
    \caption{Total galaxy stellar mass against SMBH mass for different subsets of the Horizon-AGN sample. In each panel the black points are those early-type galaxies from \protect\cite{haringrix04} with the fit shown by the black dashed line (red shaded region shows $\pm1\sigma$). In addition we show the fits from \protect\cite{greene20} to early- (red dashed line) and late-type (blue dashed line) galaxies, with the shaded regions showing $\pm1\sigma$). The top panels show galaxies which have had neither a major nor a minor merger since $z=2$, coloured by their bulge-to-total mass ratio ($B\/T$; left) and by the fraction of the SMBH mass which was built by BH mergers ($f_{\rm{BH,merge}}$; right). The bottom panels show galaxies which have had more than 3 major or minor merger since $z=2$ and are coloured by bulge-to-total ratio (left) and by the fraction of the SMBH mass which was built by BH mergers ($f_{\rm{BH,merge}}$; right). The same scales have been used on corresponding colour bars across each panel for ease of comparison. In the bottom right corner of each panel we provide the Pearson correlation coefficient, given by the $r$ value, along with the slope of the fit to the simulation data given by the $\beta$ value, and the standard deviation of the points around that fitted slope by the $sigma$ value. A value of $r=+1$ indicates a positive linear correlation, whereas a value of $r=0$ indicates no correlation. A larger value of $\beta$ indicates a steeper correlation, and a larger value of $\sigma$ indicates higher scatter around the correlation. A tighter correlation is found between SMBH mass and total stellar mass, than with bulge stellar mass, for both the \textsc{merger-free} and \textsc{merger-dominated} samples. The correlation is independent of the B$/$T ratio.}
    \label{fig:totalfig}
    \end{center}
\end{figure*}

\subsection{Velocity Dispersion Calculation}{\label{sec:veldisp}}

Horizon-AGN provides the dispersion of the stellar velocity distributions in each galaxy in Cartesian coordinates within a specified radius. In order to compare simulated velocity dispersions to observed stellar velocity dispersions measured from a galaxy spectrum (which suffers from line of sight and instrumental biases), we first extracted the velocity dispersions within a radius of $0.55R_{\rm{eff}}$, the average coverage of a 3" diameter Sloan Digital Sky Survey (SDSS) central fibre aperture at a redshift of $z=0.0556$. We then ``observed'' the velocity dispersions along a line of sight at $45^{\circ}$ to each of the Cartesian vectors to get a set of three  $\sigma_{\rm{los}, xyz}$ for each galaxy. These are then combined into a single line-of-sight velocity dispersion, $\sigma_{\rm{los}}$ as follows:
\begin{equation}\label{eq:sigmalos}
\sigma_{\rm{los}} = \sqrt{\sigma_{\rm{los, x}}^2 + \sigma_{\rm{los, y}}^2 + \sigma_{\rm{los, z}}^2} .
\end{equation}
We also account for an average instrument dispersion, $\sigma_{\rm{inst}}$, to get an equivalent observed stellar velocity dispersion, $\sigma_{\rm{obs}}$, as follows:

\begin{equation}\label{eq:obssigma}
\sigma_{\rm{obs}}^2 = \sigma_{\rm{los}}^2 + \sigma_{\rm{inst}}^2.
\end{equation}

Here we again used the instrumental dispersion of the Sloan Digital Sky Survey (SDSS; \citealt{york00}) spectrograph, $\sigma_{\rm{inst}}=69\ \rm{km}~\rm{s}^{-1}$ \citep{bolton12}. SDSS provides a large enough galaxy sample for the average fibre coverage to be estimated, and is a common data source in observational literature which we will use to discuss the context of our simulation results here.

\begin{figure*}
    \begin{center}
	\includegraphics[width=0.49\textwidth]{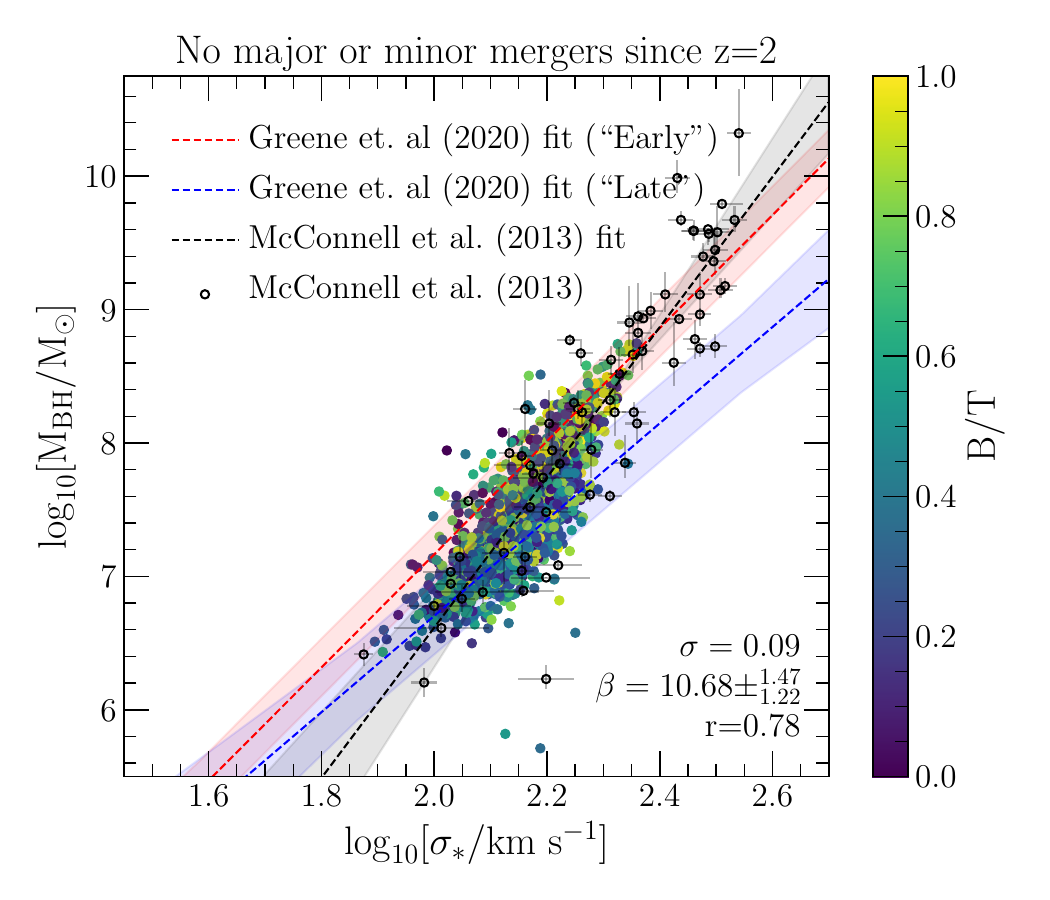}
	\includegraphics[width=0.49\textwidth]{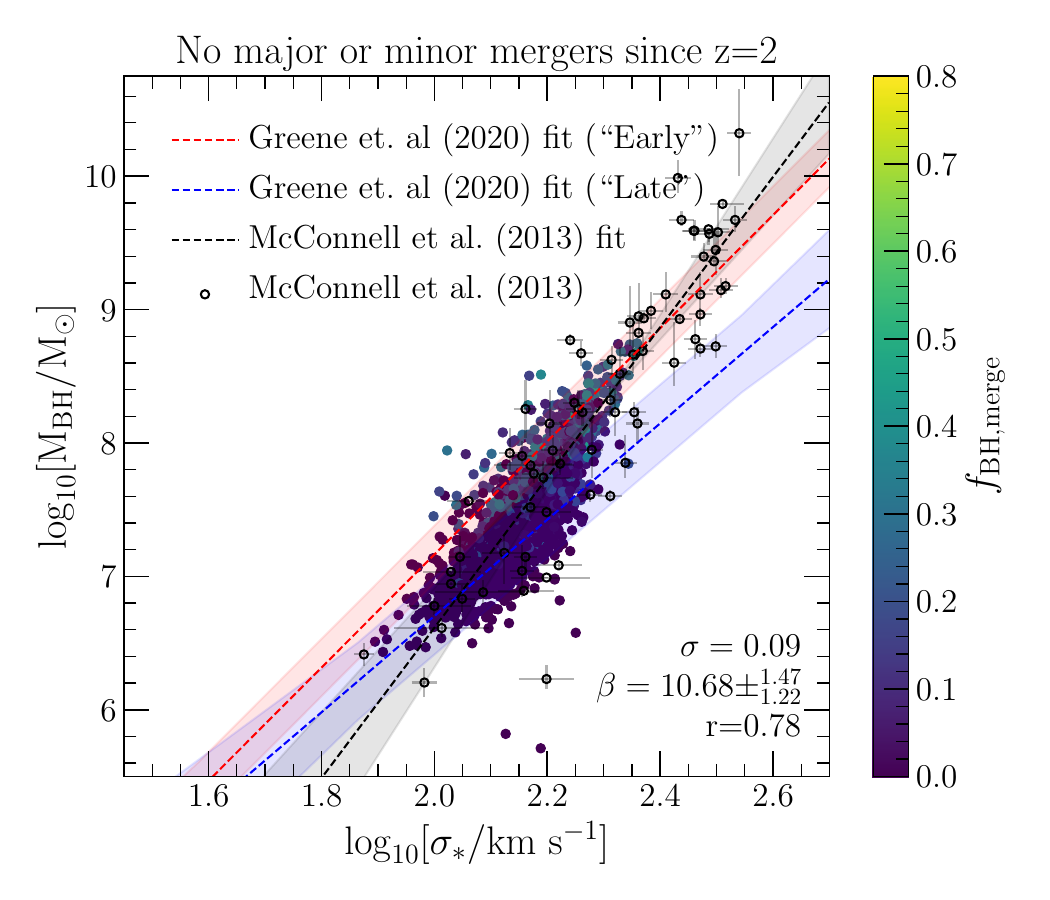}
	\includegraphics[width=0.49\textwidth]{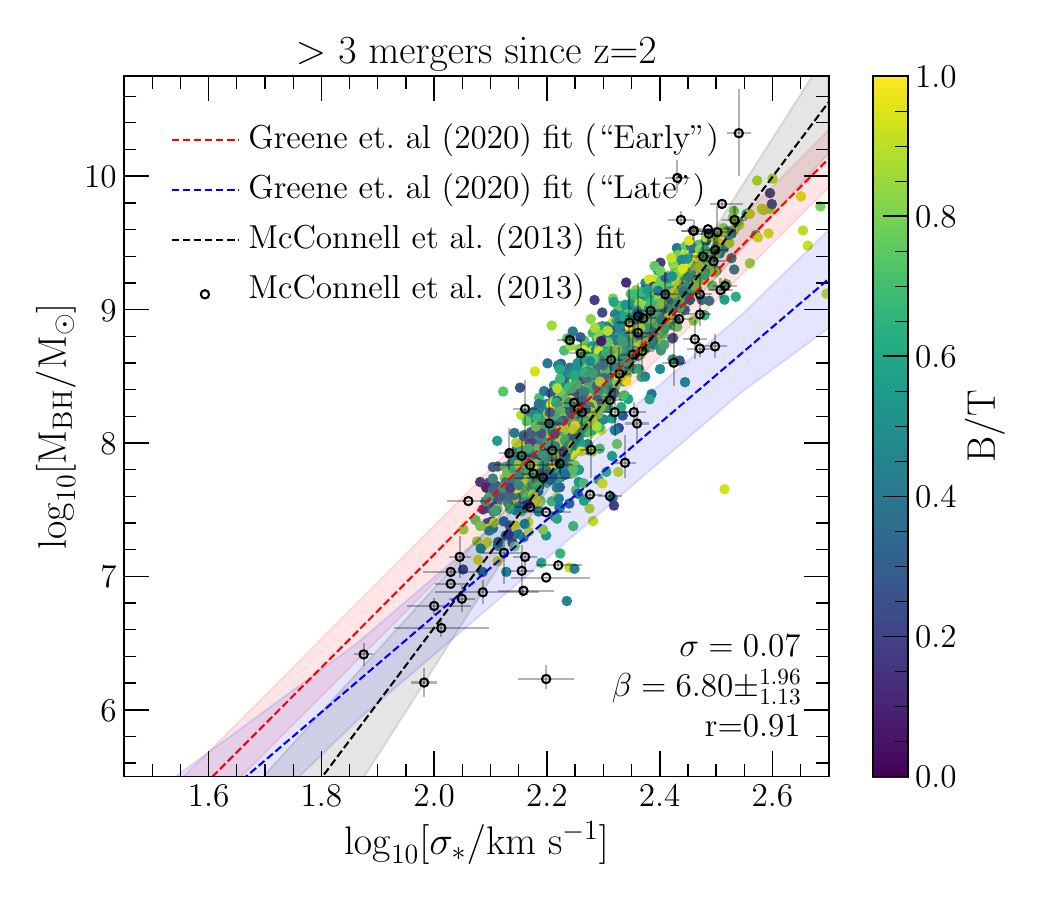}
	\includegraphics[width=0.49\textwidth]{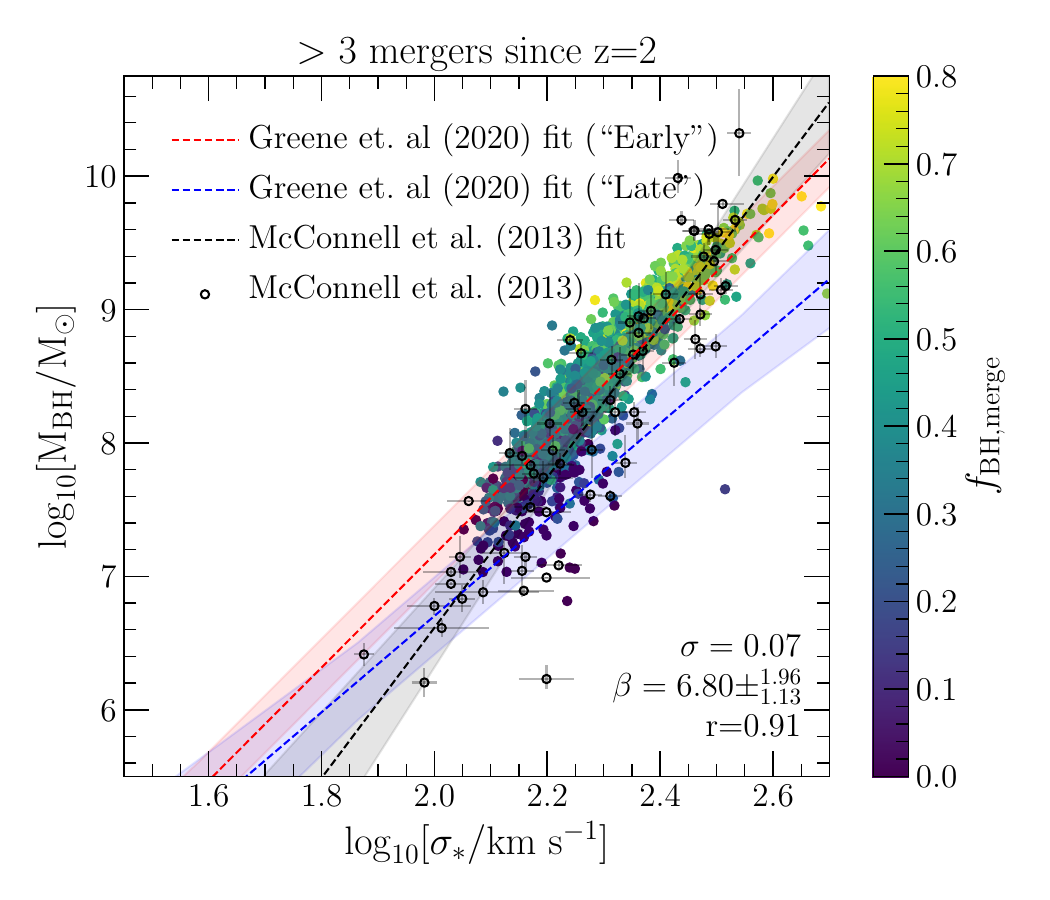}
	\vspace{-1em}
\caption{Stellar velocity dispersion against SMBH mass for different subsets of the Horizon-AGN sample. In each panel the black points are observations from \protect\cite{mcconnell13} with the fit shown by the black dashed line (red shaded region shows $\pm1\sigma$). Similarly we also show the fits from \protect\cite{greene20} for early- (red line and shaded region showing $\pm1\sigma$) and late-type (blue line and shaded region showing $\pm1\sigma$). The top panels show galaxies which have had neither a major nor a minor merger since $z=2$, coloured by their bulge-to-total mass ratio (left) and by the fraction of the SMBH mass which was built by BH mergers ($f_{\rm{BH,merge}}$; right). The bottom panels show galaxies which have had more than 3 major or minor mergers since $z=2$ and are coloured by bulge-to-total ratio (left) and by the fraction of the SMBH mass which was built by BH mergers ($f_{\rm{BH,merge}}$; right). The same scales have been used on corresponding colour bars across each panel for ease of comparison. In the bottom right corner of each panel we provide the Pearson correlation coefficient, given by the $r$ value, along with the slope of the fit to the simulation data given by the $\beta$ value, and the standard deviation of the points around that fitted slope by the $sigma$ value. A value of $r=+1$ indicates a positive linear correlation, whereas a value of $r=0$ indicates no correlation. A larger value of $\beta$ indicates a steeper correlation, and a larger value of $\sigma$ indicates higher scatter around the correlation. A tighter correlation between SMBH mass and stellar velocity dispersion is found for the \textsc{merger-dominated} sample than the \textsc{merger-free} sample, but is independent of $B\/T$ ratio. Please note the difficulties between comparing observing and simulated velocity dispersions discussed in section \ref{sec:veldisp}.}
    \label{fig:sigmafig}
    \end{center}
\end{figure*}

Even with careful work to reproduce observed velocity dispersions, there is still potential for differences between $\sigma$ values ``observed'' in simulations and those measured from on-sky astrophysical data. For example, the dispersion values in the simulations encompass dynamical information that by definition includes a rotation component, whereas dispersion values measured in a spectrum may miss substantial portions of this component depending on the orientation of the galaxy in the line of sight \citep[e.g.][]{bellovary14, vandenbosch16}. While observers generally try to correct for this and other factors \citep[e.g.][]{gultekin09,mcconnell13}, uncertainties can depend on observed properties that are non-uniform across a full galaxy population, and thus such corrections may not fully capture the quantity ``observed'' in simulations. This may lead to more pronounced differences in some galaxy populations (e.g. disk-dominated galaxies, where rotational measurements may be highly uncertain for more face-on galaxies) than others, when comparing Horizon-AGN ``observed'' dispersions and on-sky measured dispersions. Comparisons between velocity dispersions from simulations and observations must therefore always consider these potentially non-uniform biases. 

\section{Results and Discussion}\label{sec:results}

\begin{figure*}
\begin{center}
\includegraphics[width=\textwidth]{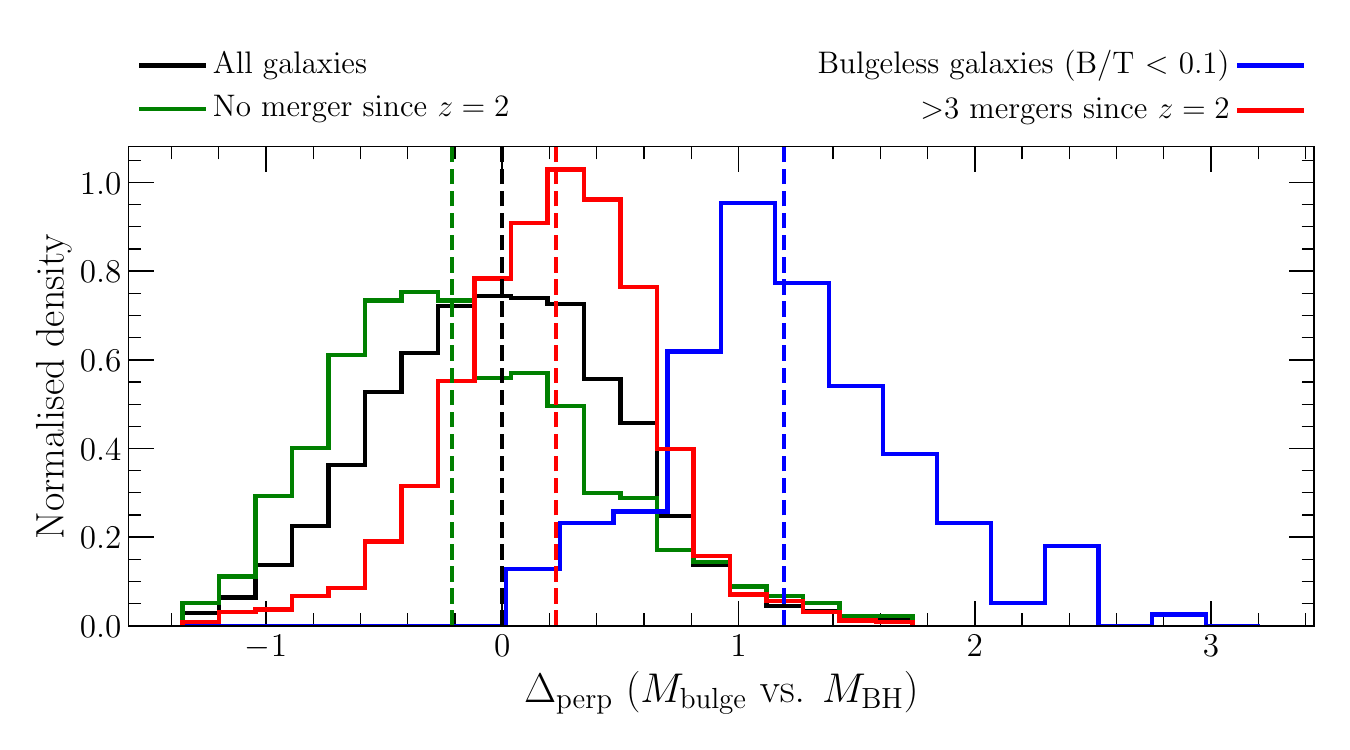}
\vspace{-1em}
\caption{Histograms showing the distribution of perpendicular distance from the best-fit relation (performed on all simulated galaxies in our sample) between bulge stellar mass and SMBH mass, $\Delta_{\rm{perp}} (M_{\rm{bulge}}$ vs. $M_{\rm BH})$ for different subsets of the Horizon-AGN sample. Dashed lines show the median value of each distribution. A negative value of $\Delta_{\rm{perp}} (M_{\rm{bulge}}$ vs. $M_{\rm BH})$ means that a galaxy lies below the relation (i.e. has a larger bulge stellar mass and lower SMBH mass than expected) and a positive value means a galaxy lies above the relation (i.e. a lower bulge stellar mass and larger SMBH mass than expected). The distribution of data points are shown in Fig. \ref{fig:bulgefig}. 
All the distributions are statistically significantly different from each other in an Anderson-Darling test \citep[$>3\sigma$;][]{anderson52} but with the \textsc{merger-free} sample (green) broader than the \textsc{merger-dominated} sample (red), reflecting the scatter seen in Fig.~\ref{fig:bulgefig}. Bulgeless galaxies (with $B/T < 0.1$; blue) are particularly anomalous, with an average shift of $\sim2.8\ \rm{dex}$ above the bulge mass-SMBH mass correlation.} 
\label{fig:bulgehist}
\end{center}
\end{figure*}

\begin{figure*}
\begin{center}
\includegraphics[width=\textwidth]{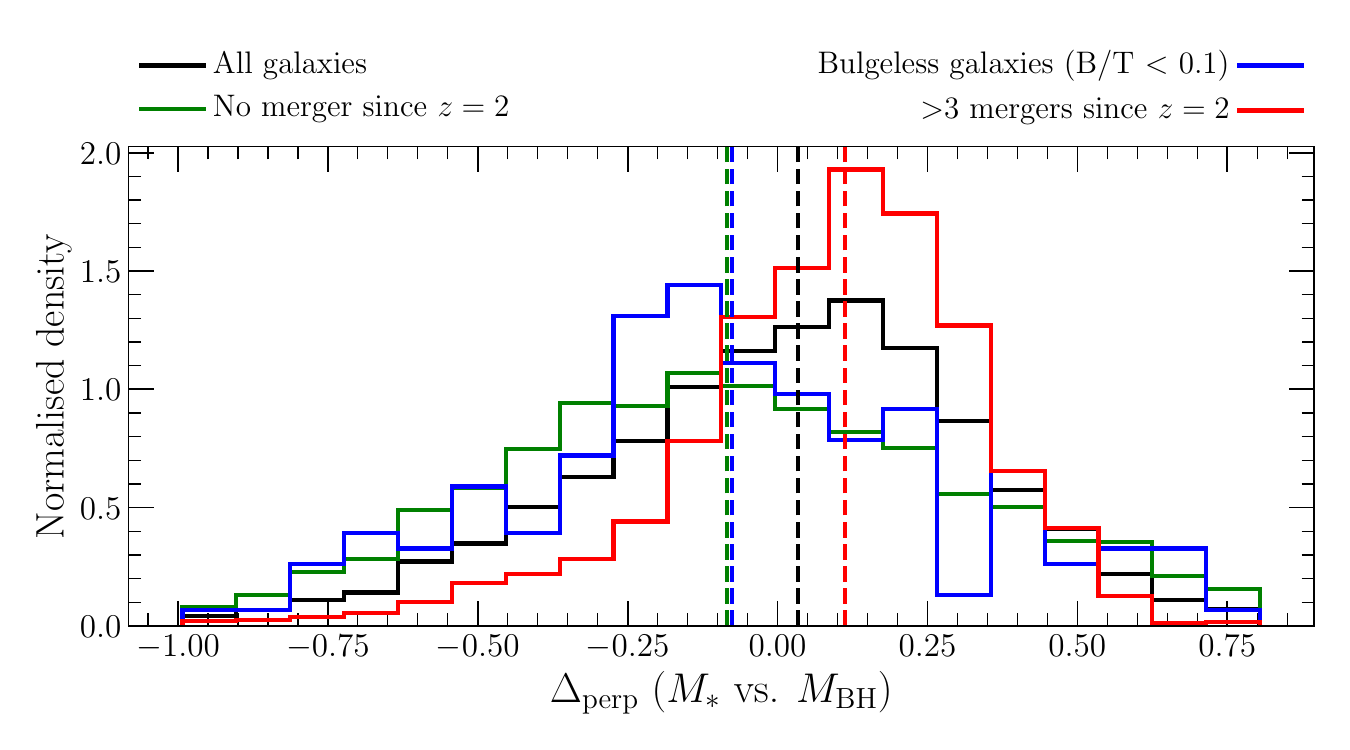}
\vspace{-1em}
\caption{Histograms showing the distribution of perpendicular distance from the best-fit relation (performed on all simulated galaxies in our sample) between total stellar mass and SMBH mass, $\Delta_{\rm{perp}} (M_*$ vs. $M_{\rm BH})$ for different subsets of the Horizon-AGN sample. Dashed lines show the median value of each distribution. A negative value of $\Delta_{\rm{perp}} (M_*$ vs. $M_{\rm BH})$ means that a galaxy lies below the relation (i.e. larger total mass and lower SMBH mass than expected) and a positive value means a galaxy lies above the relation (i.e. lower total mass and larger SMBH mass than expected). All the distributions are statistically significantly different from each other in an Anderson-Darling test \citep[$>3\sigma$;][]{anderson52}, except for the bulgeless galaxies (blue) and \textsc{merger-free} sample (green) which are statistically indistinguishable ($p=0.25$, $\sigma=1.2$). This suggests that the observationally studied bulgeless galaxies are not a uniquely evolving population, but do indeed represent the merger free population.}
\label{fig:hist}
\end{center}
\end{figure*}

\begin{figure*}
\begin{center}
\includegraphics[width=\textwidth]{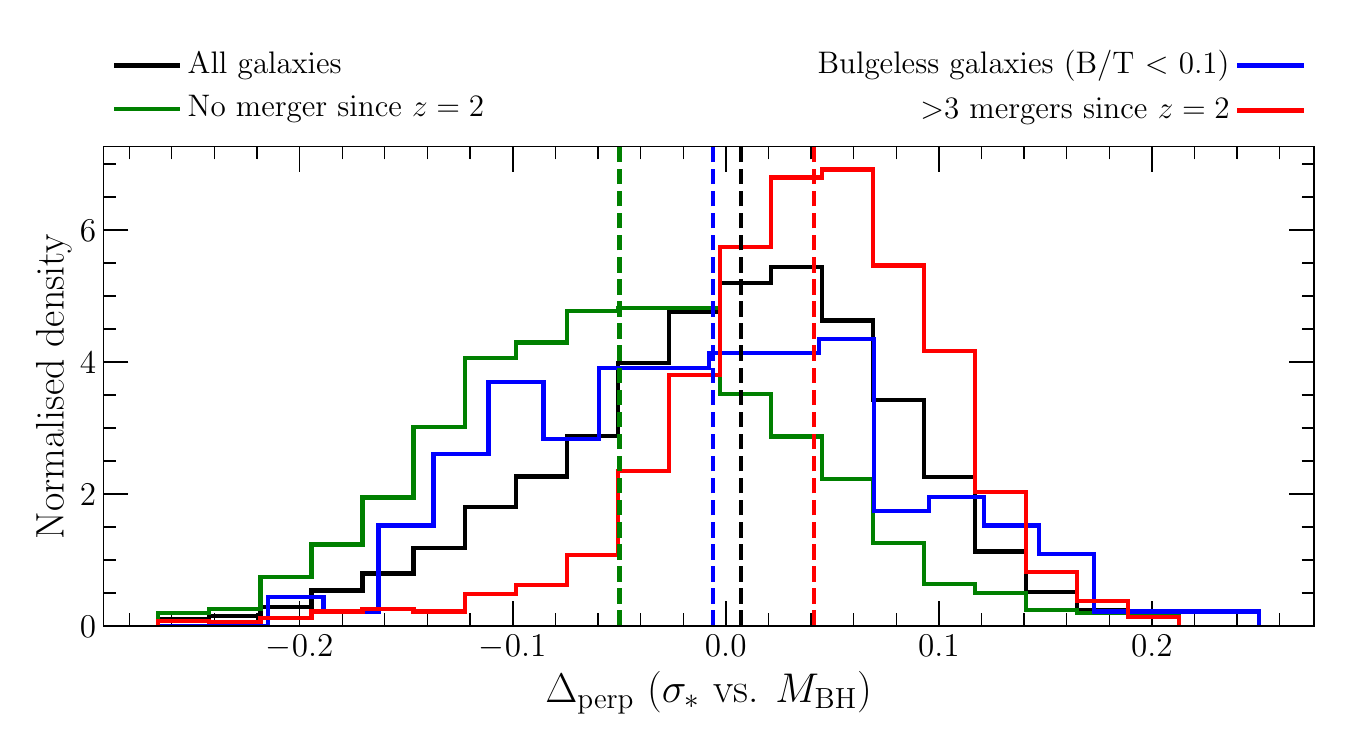}
\vspace{-1em}
\caption{Histograms showing the distribution of perpendicular distance from the best-fit $M-\sigma_*$ relation (performed on all simulated galaxies in our sample) between stellar velocity dispersion and SMBH mass, $\Delta_{\rm{perp}} (\sigma_*$ vs. $M_{\rm BH})$ for different subsets of the Horizon-AGN sample. Dashed lines show the median value of each distribution. A negative value of $\Delta_{\rm{perp}} (\sigma_*$ vs. $M_{\rm BH})$ means that a galaxy lies below the $M-\sigma_*$ relation (i.e. has a lower SMBH mass and higher $\sigma_{*}$ than expected) and a positive value means a galaxy lies above the relation (i.e. has a larger SMBH mass and lower $\sigma_{*}$ than expected). All the distributions are statistically significantly different from each other in an Anderson-Darling test \citep[$>3\sigma$;][]{anderson52}. The \textsc{merger-free} sample (green) lies below the $M-\sigma_{*}$ correlation, whereas the \textsc{merger-dominated} sample (red) lies above the correlation, as can also be seen in Fig.~\ref{fig:sigmafig}. This results in a higher intercept for a fitted $M-\sigma{*}$ relation for a merger dominated sample, and a lower fitted intercept for a merger free sample.}
\label{fig:sigmahist}
\end{center}
\end{figure*}

Figures \ref{fig:bulgefig}, \ref{fig:totalfig} \& \ref{fig:sigmafig} show the bulge stellar mass, total stellar mass and ``observed'' velocity dispersion (respectively) against the SMBH mass for the \textsc{merger-free} (top) and \textsc{merger-dominated} (bottom) samples. In the left panels the simulated galaxies are coloured by their bulge-to-total mass ratios ($B/T$) and in the right panels they are coloured by the fraction of the SMBH mass which was built by BH mergers ($f_{\rm{BH,merge}}$). We note that $f_{\rm{BH,merge}}$ is the fraction of BH mass gained through BH-BH mergers, which is not the same as the mass gained through accretion following galaxy mergers studied by \citet{mcalpine20}. In each panel of Figs. \ref{fig:bulgefig} \& \ref{fig:totalfig} we show the observations of early-type galaxies of \citet[][assumed $B/T=1$]{haringrix04} with black points, and the fit to these data points in the black dashed line ($\pm1\sigma$ is shown by the black shaded region). We performed this fit using a multiple linear regression model which encompasses the uncertainties on both x- and y-dimensions and the intrinsic scatter in the data \citep[][available as a \emph{Python}\footnote{\url{ http://linmix.readthedocs.org/}} module named \textsc{linmix}]{kelly07}. Similarly, in Fig.~\ref{fig:sigmafig} we show the observations of \citet{mcconnell13}  with black points, and the fit we made to these data points using the method above with the black dashed line.  In addition in each panel of Figs. \ref{fig:bulgefig}, \ref{fig:totalfig} \& \ref{fig:sigmafig} we plot the appropriate fits from \cite{greene20} for early- (red dashed lines and shaded regions showing $\pm1\sigma$) and late-type (blue dashed lines and shaded regions showing $\pm1\sigma$) galaxies as a comparison. We also use a multiple linear regression model to perform a fit to the simulation samples. We do not show these fits for clarity, however provide the slope of each fit, $\beta$ with $\pm1\sigma$ uncertainties, in the bottom right corner of each panel in Figs.~\ref{fig:bulgefig}, \ref{fig:totalfig} \& \ref{fig:sigmafig}. Similarly, we also provide the standard deviation of the distance of the simulated galaxies from these fits in order to quantify the scatter around the relation with the $\sigma$ value quoted in each panel, and give the Pearson correlation coefficient, $r$, which can vary from a value of $0$ for no correlation, and $+1$ for a positive linear correlation, to quantify the strength of the correlation.

Fig.~\ref{fig:bulgefig} shows that the scaling relation between bulge and SMBH mass is still present for those galaxies which have not undergone a merger, albeit with more scatter than those that have undergone a merger (demonstrated by the larger value of $\sigma$). In both the top left and bottom left panels it is clear that the bulge-to-total mass ratio ($B/T$) sets the intercept of the scaling relation, with a higher $B/T$ ratio resulting in a lower intercept for both merger driven and non-merger driven evolution. Therefore, regardless of the mechanism powering the bulge-growth (whether mergers or secular processes), and to what extent, co-evolution of the galaxy and SMBH clearly occurs. We note that due to the limited resolution of  Horizon-AGN, some secular processes such as disc instabilities and secular bar formation are most likely under-resolved. The lack of resolution acts like an extra source of  temperature in the disc which most likely prevents it from secularly barring. The results presented here should therefore be considered a lower limit as to how much galaxies and SMBHs can co-evolve through secular processes.

The bottom panels of Fig.~\ref{fig:bulgefig} show that mergers result in the most massive SMBHs and bulges, driving both more black hole and galaxy stellar mass growth. There is a clear trend with the mass of the SMBH and the fraction of the SMBH mass which was built by BH mergers ($f_{\rm{BH,merge}}$) in the bottom right panel. This is also reflected in the top right panel for the \textsc{merger-free} sample, to a lesser extent, but suggests that galaxy mergers during hierarchical structure formation at $z>2$ can lead to a more massive SMBH at low redshift in those galaxies that have since evolved in the absence of a galaxy merger.  In the Horizon-AGN simulation, only one BH can be formed per galaxy. As a result, BH mergers can only occur after galaxy mergers, and so it is expected that merger-poor galaxies also host SMBHs that have grown predominantly through accretion, just as we see in Figs. \ref{fig:bulgefig}, \ref{fig:totalfig} \& \ref{fig:sigmafig}.

In Fig.~\ref{fig:totalfig} the correlation between total stellar mass and SMBH mass is stronger and tighter than for both the bulge stellar mass (Fig.~\ref{fig:bulgefig}) and stellar velocity dispersion (Fig.~\ref{fig:sigmafig}), as demonstrated by the larger Pearson correlation co-efficient values, $r$, and smaller $\sigma$ values quoted in the bottom right corner of each panel. Once again, the \textsc{merger-dominated} sample have a stronger and tighter correlation than the \textsc{merger-free} sample. In addition, there is no dependence on the bulge-to-total ratio, $B/T$, for either those which have or haven't had mergers, as shown in the left panels. While the \textsc{merger-dominated} sample lie along the observed fit of \citet[][black dashed line]{haringrix04} and early-type fit of \citet[][red dashed line]{greene20}, the \textsc{merger-free} sample have either higher SMBH masses or lower total stellar masses than predicted by the observed late-type fit from \citet[][blue dashed line]{greene20}, and instead lie on the observed early-type relations of \citet[][red dashed line]{greene20} and \cite[][black dashed line]{haringrix04}. 

This result once again highlights the difficulties of comparing observations and simulations, and there are a number of possibilities that could explain this apparent discrepancy. Firstly, it is a well known problem that simulations struggle at replicating the scatter seen in observations across galaxy-SMBH scaling relations \citep{Habouzit2021}, which could be of particular importance for late-type galaxies which show increased scatter in stellar mass estimates using different methods \citep[for example see][]{kannappan07}. However the discrepancy in the top panels of Figure~\ref{fig:totalfig} could also be due to observational biases. For example, the galaxies in the \citet{greene20} sample are concentrated in a small local volume, with an average distance of $21.5\rm{Mpc}$ ($z=0.005$), leading to selection effects which sample SMBHs and galaxies that are known to be atypical \citep{kormendy10}. In addition, the need for simultaneous classification of both AGN activity and morphological type may lead to further selection biases which are not yet fully understood. For example \citet[][from which the \citealt{greene20} sample is constructed]{reines15} exclude systems where the AGN outshines the galaxy, which could result in lower mass SMBHs in higher mass galaxies. Upon inspection, the late-type fits of \cite{greene20} include very few galaxies under $10^{10}~\rm{M}_{\odot}$ where the discrepancy in Figure~\ref{fig:totalfig} is most apparent. There is also a debate in the literature whether there is truly an offset in the scaling relations for different morphological types \citep[e.g.][]{salucci00, bentz18, sahu19, davis19, bennert21}. The left hand panels of Figure~\ref{fig:totalfig} suggest that when morphology is probed using bulge-to-total mass ratio, $B/T$, as a proxy, there is no offset between scaling relations, instead an increase in $B/T$ will reduce the scatter around the relation.

In Fig.~\ref{fig:sigmafig}, the \textsc{merger-dominated} sample have both higher observed velocity dispersions and SMBH masses than the \textsc{merger-free} sample (as expected). Once again, there is no correlation with galaxy $B/T$ ratio for either sample, however the slope of the $M_{\rm BH}-\sigma_{*}$ relation appears to be set by the fraction of the SMBH mass grown by BH mergers, $f_{\rm{BH,merge}}$ (see right panels of Fig.~\ref{fig:sigmafig}). The \textsc{merger-free} sample still appear to lie on the $M-\sigma_*$ correlations fit to the observations of \citet[][shown by the black dashed line]{mcconnell13} and late-type observations of \citet[][shown by the blue dashed line]{greene20}, albeit with a larger scatter. However, it is intriguing to note that the \textsc{merger-dominated} sample mostly lie above the $M_{\rm BH}-\sigma_{*}$ observed correlations of \citet{mcconnell13} and \citet{greene20}, with larger SMBH masses than expected given their velocity dispersions. This could be due to the the difficulties of directly comparing simulated and observed velocity dispersions due to the non-uniform biases present in dispersions measured from spectra (see Section~\ref{sec:veldisp}). 

Fig.~\ref{fig:bulgehist} shows the perpendicular distance of galaxies in the Horizon-AGN sample from the best-fit relation between bulge stellar mass and SMBH mass, $\Delta_{\rm{perp}} (M_{\rm{bulge}}$ vs. $M_{\rm BH})$. A negative value of $\Delta_{\rm{perp}} (M_{\rm{bulge}}$ vs. $M_{\rm BH})$ means that a galaxy lies below the relation (i.e. has a larger bulge mass and lower SMBH mass than expected) and a positive value means a galaxy lies above the relation (i.e. a lower bulge mass and larger SMBH mass than expected). All the histograms are statistically significantly different from each other (Anderson-Darling test $\sigma>3.3$; \citealt{anderson52}), with the \textsc{merger-free} sample having a broader distribution than the \textsc{merger-dominated} sample, reflecting the larger scatter seen in the top panels of Fig.~\ref{fig:bulgefig}. The `bulgeless' galaxies with $B/T <0.1$ are particularly anomalous, with perpendicular distances of $\sim2.8~\rm{dex}$ above the fitted relation, in agreement with observations \citep{ssl17} and previous work with Horizon-AGN \citep{martin18}. Recall that the discrepency is stronger at the low mass end than at the high mass end (see Fig. \ref{fig:bulgefig}). The SMBHs of these `bulgeless' galaxies are more massive than expected given their lack of bulge however, as Fig.~\ref{fig:hist} shows, not for their total stellar mass. This once again clearly shows that SMBH growth can occur in the absence of bulge growth.

Similarly, Fig.~\ref{fig:hist} shows the distribution of the perpendicular distance of galaxies in the Horizon-AGN sample from the best-fit relation between total stellar mass and SMBH mass, $\Delta_{\rm{perp}} (M_*$ vs. $M_{\rm BH})$. A negative value of $\Delta_{\rm{perp}} (M_*$ vs. $M_{\rm BH})$ means that a galaxy lies below the relation (i.e. has a larger stellar mass and lower SMBH mass than expected) and a positive value means a galaxy lies above the relation (i.e. a larger SMBH mass and lower total stellar mass than expected). The \textsc{merger-dominated} sample have statistically significantly ($>3\sigma$) higher values of $\Delta_{\rm{perp}} (M_*$ vs. $M_{\rm BH})$ and therefore larger SMBH masses than expected from the typical scaling relation. Conversely the \textsc{merger-free} sample have statistically significantly ($>3\sigma$) lower $\Delta_{\rm{perp}} (M_*$ vs. $M_{\rm BH})$ values and therefore larger stellar masses than expected. In addition the `bulgeless' galaxies, selected as $B/T<0.1$, which are studied observationally as a way to isolate merger-free SMBH growth also have lower $\Delta_{\rm{perp}} (M_*$ vs. $M_{\rm BH})$ values, and are statistically indistinguishable ($p=0.25$, $\sigma=1.2$) from galaxies which have not had a merger since $z=2$ in the Horizon-AGN sample. This suggests that the `bulgeless' galaxies studied observationally by \citet{simmons13}, \citet{ssl17}, \citet{smethurst19} and \citet{smethurst21} are not a uniquely evolving population, but do indeed represent the merger-free growth pathway occurring across the galaxy population.

Fig.~\ref{fig:sigmahist} shows the distribution of the perpendicular distance of galaxies in the Horizon-AGN sample from the best-fit relation between stellar velocity dispersion and SMBH mass, $\Delta_{\rm{perp}} (\sigma_*$ vs. $M_{\rm BH})$. A negative value of $\Delta_{\rm{perp}} (\sigma_*$ vs. $M_{\rm BH})$ means that a galaxy lies below the relation (i.e. has a larger velocity dispersion and lower SMBH mass than expected) and a positive value means a galaxy lies above the relation (i.e. a larger SMBH mass and lower velocity dispersion than expected). The \textsc{merger-dominated} sample have statistically significantly \mbox{($>3\sigma$)} higher values of $\Delta_{\rm{perp}} (\sigma_*$ vs. $M_{\rm BH})$ and therefore larger SMBH masses than expected from the typical scaling relation. Conversely the \textsc{merger-free} sample have statistically significantly ($>3\sigma$) lower $\Delta_{\rm{perp}} (\sigma_*$ vs. $M_{\rm BH})$ values and therefore larger velocity dispersions than expected. In addition the `bulgeless' galaxies, selected as $B/T<0.1$, which are studied observationally as a way to isolate merger-free SMBH growth have a broad range in $\Delta_{\rm{perp}} (\sigma_*$ vs. $M_{\rm BH})$ values, similar to that found for the entire sample of all morphologies. The distributions in Fig.~\ref{fig:sigmahist} support the results of \cite{bell17} who found observational evidence for `classical' bulge growth (i.e. increased velocity dispersions) in the absence of mergers.

Assuming galaxies from the Horizon-AGN simulation are reflective of the galaxy population, it is interesting to consider that $26\%$ of galaxies have not undergone a galaxy merger since $z=2$, as opposed to only $18\%$ which have undergone more than 3 mergers. The other $56\%$ of the galaxy population is therefore evolving through a mix of galaxy mergers and non-merger processes. Even for galaxies that repeatedly experience mergers, there are long periods of time between mergers when galaxies evolve secularly. Given previous observational and simulated results, this suggests non-merger processes play an important role in galaxy-SMBH co-evolution for a wide range of different galaxy evolution histories. 

Co-evolution is thought to be regulated by AGN feedback. \cite{smethurst21} investigated the ionised outflows from \rjs{ 4 optically selected AGN in `bulgeless' systems using the Keck Cosmic Web Imager (KCWI; an IFU at the Keck Observatory)}, finding that the outflows have velocities far exceeding the escape velocity of their galaxies and extend over kiloparsec scales ($0.6-2.4~\rm{kpc}$), suggesting that may be capable of causing feedback. \rjs{Similarly, \cite{bohn22} found that 5 of their 9 infrared selected AGN hosted by `bulgeless' galaxies had  AGN powered outflows in either [OIII] and/or [Si VI]. In addition \citeauthor{bohn22} investigated the rotational gas kinematics of their sample and found 2 of their `bulgeless' AGN hosts were rotating faster than expected, resulting in lower stellar masses than expected from abundance matching. \citeauthor{bohn22} interpret this as AGN feedback having suppressed star formation in these 2 `bulgeless' merger-free galaxies leading to the reduction in stellar mass.  The studies of both \cite{smethurst21} and \cite{bohn22}} combined with our results in this work have interesting implications: if both merger-driven and non-merger-driven growth can lead to galaxy-SMBH co-evolution, this suggests that co-evolution could be regulated by AGN feedback in both scenarios.

One noticeable feature of  Figs.~\ref{fig:bulgefig}, \ref{fig:totalfig} \& \ref{fig:sigmafig} is that the correlation between bulge or total stellar mass and SMBH mass for the \textsc{merger-dominated} sample is tighter, with less scatter (see standard deviation values, $\sigma$, given in each panel) than for the \textsc{merger-free} sample. This is likely due to the averaging effect of galaxy mergers: statistically, a merger is likely to include a galaxy on either side of the mean, so adding their masses implies that the merger remnant is likely to lie closer to the mean than either of the progenitors \cite{Knud2011}. It is a well known problem that galaxy evolution simulations tend to lack scatter in the $M_{\rm BH} - M_*$ relation in comparison to observations \citep{Habouzit2021}. It is therefore intriguing to see here that the scatter that does exist is dominated by merger-free systems, with the reduction in scatter caused by mergers and associated processes. From our sample, we predict that the scatter for secularly evolved galaxies should be much larger than for those undergoing repeated mergers. This also means that galaxies at higher redshift would be expected to have larger scatter around typical scaling relations than those at low redshift.

\begin{figure}
    \centering
    \includegraphics[width=\columnwidth]{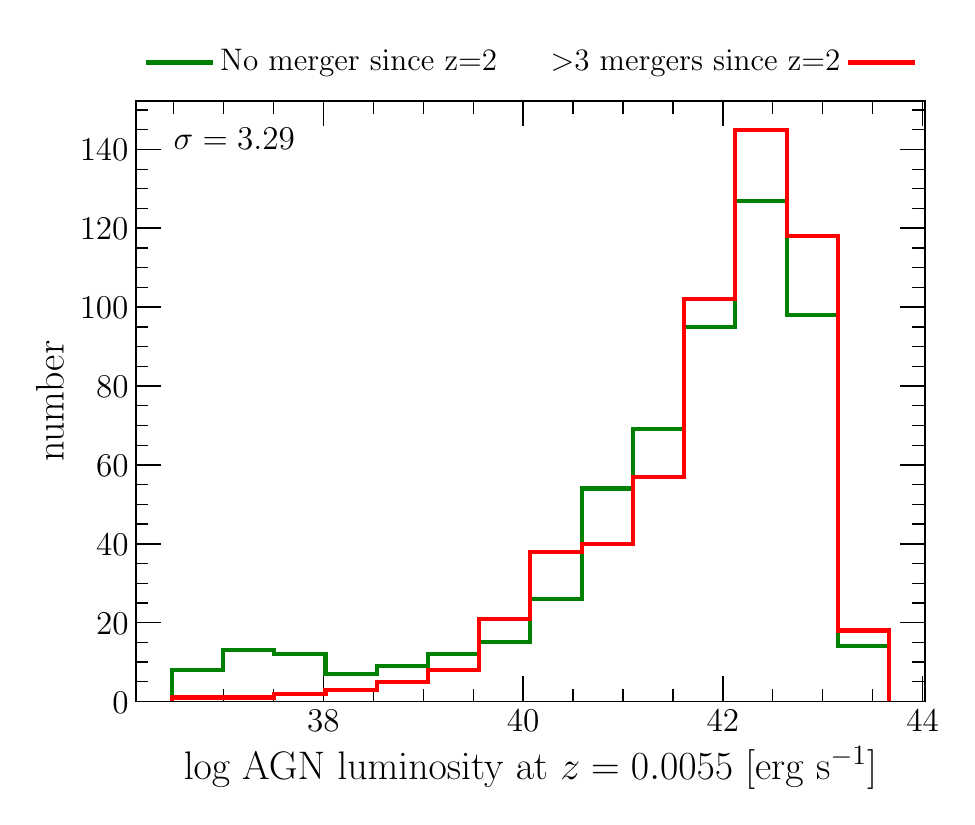}
    \caption{Distribution of AGN luminosity at $z=0.00556$ for the \textsc{merger-free} and \textsc{merger-dominated} samples matched in SMBH mass, resulting in 563 galaxies. The two distributions are statistically indistinguishable from each other in an Anderson-Darling test \citep[$>3\sigma$;][]{anderson52}.}
    \label{fig:LAGN}
\end{figure}

The intriguing question of what drives the secular co-evolution between SMBH and their merger-free host galaxies demonstrated in this paper, remains open. One likely candidate is AGN feedback. As discussed in the companion paper to this work, \citet{beckmann22}, AGN in \textsc{merger-free} and \textsc{merger-dominated} galaxies look remarkably similar in some ways. In Figure~\ref{fig:LAGN} we find that after adjusting for the difference in the SMBH mass between the two simulated galaxy samples, at $z=0.00556$, there is no statistically significant difference ($>3\sigma$) in the distribution of luminosities between AGN hosted in the \textsc{merger-dominated} and \textsc{merger-free} galaxies. \rb{This is despite the well-established fact that galaxy mergers cause spikes in AGN luminosity \citep[see e.g.][among others]{gabor15,steinborn18,mcalpine20}, a phenomenon also exhibited by AGN in the Horizon-AGN simulation \citep[][]{volonteri16}. However, at any given redshift the incidence of on-going galaxy mergers in the sample is small, so at $z=0.00556$ the majority of galaxies even in the \textsc{merger-dominated} sample will also currently be co-evolving secularly with their host galaxy.}

\rb{To understand how the impact of AGN feedback might differ for \textsc{merger-dominated} and \textsc{merger-free} galaxies, we need to study the evolution of AGN  feedback over time, not just instantaneously. In Horizon-AGN we model two modes of feedback, the `jet/radio' mode, in which AGN energy is delivered as kinetic energy, and the `quasar/thermal' mode, in which it is delivered as thermal energy. Which mode a given AGN is in depends on the mass of the SMBH and its luminosity. Horizon-AGN was one of the first simulations to employ bimodal SMBH feedback, using a thermal or kinetic energy injection depending on the SMBH accretion efficiency. This two-mode model has become well established in the galaxy evolution community, and is now  commonly employed in large-scale galaxy evolution simulations including IllustrisTNG \citep{Weinberger2017} and SIMBA \citep{Dave2019}. As the feedback mechanisms are similar across simulations, we expect the trends reported here to also hold in other comparable datasets.}

\rb{When comparing the distribution of feedback modes over time, we again find many similarities between the two samples:} AGN in the \textsc{merger-dominated} sample spend just 3\% more of their time in `quasar/thermal' mode since $z=2$ in comparison to the \textsc{merger-free} sample. This means that for all galaxies seen here, AGN spend the majority of their time ($\sim 90 \%$ for both samples) in a `jet/radio' mode, which could suggest that such `jet/radio' feedback could play an important role in regulating star formation in the host galaxy, and in driving the co-evolution between SMBH and their host galaxies, as discussed in \citet{dave19}.

On the other hand, as AGN feedback in the `quasar/thermal' mode is powered by higher accretion rates, similar amounts of time do not necessarily translate to similar amounts of energy delivered in each mode. Here, the luminosity spikes post galaxy merger show their significance: \textsc{merger-dominated} galaxies receive 45\% of their AGN feedback energy in `quasar/thermal' mode, versus 55\% in `radio/jet' mode, so both modes are potentially equally important in driving the co-evolution between galaxies. By contrast, \textsc{merger-free} galaxies receive only 17\% of their AGN energy in `quasar/thermal' mode, compared to 83\% in `radio/jet' mode. \rb{While there is evidence that even short bursts of `quasar/thermal' mode can play an important role in driving the co-evolution between SMBH and galaxies, \rb{including in Horizon-AGN \citep{dubois16} and other large-scale cosmological simulations \citep[such as][]{bellovary14,sijacki15,steinborn18,mcalpine20}}, this hints at the intriguing possibility that the secular co-evolution in \textsc{merger-free} galaxies is driven by radio mode feedback.}

However, the impact of AGN energy on the host galaxy is not purely a function of the total amount of energy injected: occasional strong AGN bursts are more direct at quenching central star formation than constant low-level AGN luminosity leading to a radio mode of feedback, even if the total energy budget is the same. This could account for the lack of evidence for negative AGN feedback across the low-z galaxy samples selected from galaxy surveys, where a significant fraction of AGN are found in disk galaxies \citep{kauffmann03, schawinski10a, koss11, povic12, villforth14, smethurst16, RW18, zhao21, zhong22} which are dominated by non-merger co-evolution. \rjs{Future observational work with a high resolution IFU instrument supported by adaptive optics (e.g. such as MUSE on the VLT, or VIRUS on the HET) will be capable of detecting the broadened outflows specifically in a sample of merger-free systems, selected observationally as those which have `bulgeless' morphologies. Such high resolution IFU observations can resolve the morphology of the AGN outflows and probe the resolved narrow H$\alpha$ \& H$\beta$ emission ionised by star formation in the regions incident with the outflow to determine the impact of the outflows on the star formation rate (SFR) through feedback. These measurements of SFR can be compared against the predictions from simulations of the effect of AGN feedback on the SFR of a merger-free/isolated galaxy \citep[e.g.][]{barai14, dubois16, beckmann17, dave19, torrey20}. Such a sample will allow the effects of merger-free powered AGN feedback to be isolated observationally for the first time.} 

In this paper, we provided strong evidence that SMBH and their host galaxies co-evolve even in the absence of mergers, and showed that during the long periods between galaxy mergers, the AGN populations in both samples look remarkably similar. This could provide an intriguing hint to the possibility that the co-evolution between SMBH and galaxies even in galaxies that do experience frequent mergers might be at least in part driven by the long secular evolution epochs between mergers. Whether this secular co-evolution is regulated through `jet/radio' AGN feedback, or also through `thermal/quasar' feedback like the merger-driven co-evolution, and what the relative importance of merger-driven and secular evolution is for those galaxies that do frequently merge remains to be determined.

\section{Conclusions}\label{sec:conclusions}

Here we have used the Horizon-AGN simulation to isolate merger-free co-evolution of galaxies and their SMBHs. Although galaxy evolution will inevitably be a mix of both mergers and internal secular evolution, our results show that secular evolution alone over the past $11$ billion years  still results in co-evolution of galaxies and SMBHs. Our findings are summarised as follows: 

\begin{enumerate}
    \item{Correlations between SMBH mass and total stellar mass, bulge mass, and stellar velocity dispersion persist for merger-free galaxies, suggesting that co-evolution occurs in the absence of galaxy mergers.}
    \item Galaxy mergers reduce the scatter in the SMBH-galaxy scaling relations. In addition they make the correlations stronger, with a steeper slope. This is most apparent for the correlation between total stellar mass and SMBH mass (see Figure~\ref{fig:totalfig}) and our results support the hypothesis that there is no offset between the scaling relation of different morphological types. 
    \item For merger-free objects the bulge-to-total ratio, $B/T$, sets the normalisation of the $M_{\rm BH} - M_{\rm bulge} $ relation, but has no impact on the $M_{\rm BH}-M_{\rm *}$ relation. Co-evolution appears to be independent of bulge mass in both the merger and merger-free scenarios. 
    \item{Merger-free galaxies still follow the $M_{\rm BH}-\sigma_{*}$ relation, which is once again bulge mass independent.}
    \item AGN properties for SMBH in \textsc{merger-dominated} and \textsc{merger-free} galaxies look remarkably similar, suggesting that even galaxies that merge frequently might experience significant secular co-evolution with their SMBH during the long epochs between galaxy mergers.
    \item It remains to be determined if this secular co-evolution is driven by `jet/radio' mode, where AGN in both types of galaxies spend around 90\% of their time or by `quasar/thermal' mode, which provides only 17\% of total AGN energy for merger-free galaxies, as opposed to 45\% of the total AGN energy for merger-dominated galaxies.
\end{enumerate}

 Future observational work to investigate the direct and indirect impact of AGN outflows on the star formation rate in a merger-free galaxy sample, for example with a high resolution IFU instrument such as MUSE on the VLT, is therefore imperative to understand the effects of merger-free powered AGN feedback which is now thought to be the dominant mechanism regulating galaxy-SMBH co-evolution. 

\section*{Acknowledgements}
First authorship is shared between RJS and RSB. RJS conceived of the project, analysed data, interpreted results and wrote the manuscript. RSB assembled data catalogues, interpreted results and wrote the manuscript.

RJS gratefully acknowledges funding from Christ Church, Oxford and the Royal Astronomical Society. RSB gratefully acknowledges funding from Newnham College, Cambridge. BDS acknowledges support from a UK Research and Innovation Future Leaders Fellowship [grant number MR/T044136/1]. This work has made use of the Horizon Cluster hosted by Institut d’Astrophysique de Paris. We thank Stéphane Rouberol for smoothly running this cluster for us. ILG acknowledges support from an STFC PhD studentship [grant number ST/T506205/1] and from the Faculty of Science and Technology at Lancaster University.

This work used the HPC resources of CINES (Jade supercomputer) under the allocation 2013047012 made by GENCI, and the horizon and Dirac clusters for post processing. This work is partially supported by the Spin(e) grants ANR-13-BS05-0002 of the French Agence Nationale de la Recherche and by the National Science Foundation under Grant No. NSF PHY11- 25915, and it is part of the Horizon-UK project, which used the DiRAC Complexity sys- tem, operated by the University of Leicester IT Services, which forms part of the STFC DiRAC HPC Facility (www.dirac.ac.uk). This equipment is funded by BIS National E-Infrastructure cap- ital grant ST/K000373/1 and STFC DiRAC Operations grant ST/K0003259/1. DiRAC is part of the National E-Infrastructure. 

This research made use of Astropy,\footnote{http://www.astropy.org} a community-developed core Python package for Astronomy \citep{astropy13, astropy18}.

\section*{Data Availability}

All data used in this paper is available upon request to the first authors.
 



\bibliographystyle{mnras}
\bibliography{refs.bib}

\newcommand{\noop}[1]{}
\begin{thebibliography}{}
\makeatletter
\relax
\def\mn@urlcharsother{\let\do\@makeother \do\$\do\&\do\#\do\^\do\_\do\%\do\~}
\def\mn@doi{\begingroup\mn@urlcharsother \@ifnextchar [ {\mn@doi@}
  {\mn@doi@[]}}
\def\mn@doi@[#1]#2{\def\@tempa{#1}\ifx\@tempa\@empty \href
  {http://dx.doi.org/#2} {doi:#2}\else \href {http://dx.doi.org/#2} {#1}\fi
  \endgroup}
\def\mn@eprint#1#2{\mn@eprint@#1:#2::\@nil}
\def\mn@eprint@arXiv#1{\href {http://arxiv.org/abs/#1} {{\tt arXiv:#1}}}
\def\mn@eprint@dblp#1{\href {http://dblp.uni-trier.de/rec/bibtex/#1.xml}
  {dblp:#1}}
\def\mn@eprint@#1:#2:#3:#4\@nil{\def\@tempa {#1}\def\@tempb {#2}\def\@tempc
  {#3}\ifx \@tempc \@empty \let \@tempc \@tempb \let \@tempb \@tempa \fi \ifx
  \@tempb \@empty \def\@tempb {arXiv}\fi \@ifundefined
  {mn@eprint@\@tempb}{\@tempb:\@tempc}{\expandafter \expandafter \csname
  mn@eprint@\@tempb\endcsname \expandafter{\@tempc}}}

\bibitem[\protect\citeauthoryear{Anderson \& Darling}{Anderson \&
  Darling}{1952}]{anderson52}
Anderson T.~W.,  Darling D.~A.,  1952, \mn@doi [Ann. Math. Statist.]
  {10.1214/aoms/1177729437}, 23, 193

\bibitem[\protect\citeauthoryear{{Astropy Collaboration} et~al.,}{{Astropy
  Collaboration} et~al.}{2013}]{astropy13}
{Astropy Collaboration} et~al., 2013, \mn@doi [\aap]
  {10.1051/0004-6361/201322068}, \href
  {http://adsabs.harvard.edu/abs/2013A%26A...558A..33A} {558, A33}

\bibitem[\protect\citeauthoryear{{Astropy Collaboration}, {Price-Whelan},
  {Sip{\H o}cz}, {G{\"u}nther}, {Lim}, {Crawford}  \& {Astropy
  Contributors}}{{Astropy Collaboration} et~al.}{2018}]{astropy18}
{Astropy Collaboration} {Price-Whelan} A.~M.,  {Sip{\H o}cz} B.~M.,
  {G{\"u}nther} H.~M.,  {Lim} P.~L.,  {Crawford} S.~M.,   {Astropy
  Contributors} 2018, \mn@doi [\aj] {10.3847/1538-3881/aabc4f}, \href
  {http://adsabs.harvard.edu/abs/2018AJ....156..123A} {156, 123}

\bibitem[\protect\citeauthoryear{Aubert, Pichon  \& Colombi}{Aubert
  et~al.}{2004}]{aubert04}
Aubert D.,  Pichon C.,   Colombi S.,  2004, \mn@doi [\mnras]
  {10.1111/j.1365-2966.2004.07883.x}, 352, 376

\bibitem[\protect\citeauthoryear{{Baldassare}, {Dickey}, {Geha}  \&
  {Reines}}{{Baldassare} et~al.}{2020}]{baldassare20}
{Baldassare} V.~F.,  {Dickey} C.,  {Geha} M.,   {Reines} A.~E.,  2020, \mn@doi
  [\apjl] {10.3847/2041-8213/aba0c1}, \href
  {https://ui.adsabs.harvard.edu/abs/2020ApJ...898L...3B} {898, L3}

\bibitem[\protect\citeauthoryear{{Barai}, {Viel}, {Murante}, {Gaspari}  \&
  {Borgani}}{{Barai} et~al.}{2014}]{barai14}
{Barai} P.,  {Viel} M.,  {Murante} G.,  {Gaspari} M.,   {Borgani} S.,  2014,
  \mn@doi [\mnras] {10.1093/mnras/stt1977}, \href
  {https://ui.adsabs.harvard.edu/abs/2014MNRAS.437.1456B} {437, 1456}

\bibitem[\protect\citeauthoryear{{Batiste}, {Bentz}, {Raimundo}, {Vestergaard}
  \& {Onken}}{{Batiste} et~al.}{2017}]{batiste17}
{Batiste} M.,  {Bentz} M.~C.,  {Raimundo} S.~I.,  {Vestergaard} M.,   {Onken}
  C.~A.,  2017, \mn@doi [\apjl] {10.3847/2041-8213/aa6571}, \href
  {https://ui.adsabs.harvard.edu/abs/2017ApJ...838L..10B} {838, L10}

\bibitem[\protect\citeauthoryear{{Beckmann} \& {Smethurst}}{{Beckmann} \&
  {Smethurst}}{in prep}]{beckmann22}
{Beckmann} R.~S.,  {Smethurst} R.~J.,  \noop{2022}in prep, \mnras

\bibitem[\protect\citeauthoryear{{Beckmann} et~al.,}{{Beckmann}
  et~al.}{2017}]{beckmann17}
{Beckmann} R.~S.,  et~al., 2017, \mn@doi [\mnras] {10.1093/mnras/stx1831},
  \href {https://ui.adsabs.harvard.edu/abs/2017MNRAS.472..949B} {472, 949}

\bibitem[\protect\citeauthoryear{{Bell}, {Monachesi}, {Harmsen}, {de Jong},
  {Bailin}, {Radburn-Smith}, {D'Souza}  \& {Holwerda}}{{Bell}
  et~al.}{2017}]{bell17}
{Bell} E.~F.,  {Monachesi} A.,  {Harmsen} B.,  {de Jong} R.~S.,  {Bailin} J.,
  {Radburn-Smith} D.~J.,  {D'Souza} R.,   {Holwerda} B.~W.,  2017, \mn@doi
  [\apjl] {10.3847/2041-8213/aa6158}, \href
  {https://ui.adsabs.harvard.edu/abs/2017ApJ...837L...8B} {837, L8}

\bibitem[\protect\citeauthoryear{{Bellovary}, {Brooks}, {Volonteri},
  {Governato}, {Quinn}  \& {Wadsley}}{{Bellovary} et~al.}{2013}]{bellovary13}
{Bellovary} J.,  {Brooks} A.,  {Volonteri} M.,  {Governato} F.,  {Quinn} T.,
  {Wadsley} J.,  2013, \mn@doi [\apj] {10.1088/0004-637X/779/2/136}, \href
  {http://adsabs.harvard.edu/abs/2013ApJ...779..136B} {779, 136}

\bibitem[\protect\citeauthoryear{{Bellovary}, {Holley-Bockelmann},
  {G{\"u}ltekin}, {Christensen}, {Governato}, {Brooks}, {Loebman}  \&
  {Munshi}}{{Bellovary} et~al.}{2014}]{bellovary14}
{Bellovary} J.~M.,  {Holley-Bockelmann} K.,  {G{\"u}ltekin} K.,  {Christensen}
  C.~R.,  {Governato} F.,  {Brooks} A.~M.,  {Loebman} S.,   {Munshi} F.,  2014,
  \mn@doi [\mnras] {10.1093/mnras/stu1958}, \href
  {https://ui.adsabs.harvard.edu/abs/2014MNRAS.445.2667B} {445, 2667}

\bibitem[\protect\citeauthoryear{{Bennert} et~al.,}{{Bennert}
  et~al.}{2021}]{bennert21}
{Bennert} V.~N.,  et~al., 2021, \mn@doi [\apj] {10.3847/1538-4357/ac151a},
  \href {https://ui.adsabs.harvard.edu/abs/2021ApJ...921...36B} {921, 36}

\bibitem[\protect\citeauthoryear{{Bentz} \& {Manne-Nicholas}}{{Bentz} \&
  {Manne-Nicholas}}{2018}]{bentz18}
{Bentz} M.~C.,  {Manne-Nicholas} E.,  2018, \mn@doi [\apj]
  {10.3847/1538-4357/aad808}, \href
  {https://ui.adsabs.harvard.edu/abs/2018ApJ...864..146B} {864, 146}

\bibitem[\protect\citeauthoryear{{Bohn}, {Canalizo}, {Satyapal}  \&
  {Sales}}{{Bohn} et~al.}{2022}]{bohn22}
{Bohn} T.,  {Canalizo} G.,  {Satyapal} S.,   {Sales} L.~V.,  2022, \mn@doi
  [\apj] {10.3847/1538-4357/ac6870}, \href
  {https://ui.adsabs.harvard.edu/abs/2022ApJ...931...69B} {931, 69}

\bibitem[\protect\citeauthoryear{{Bolton} et~al.,}{{Bolton}
  et~al.}{2012}]{bolton12}
{Bolton} A.~S.,  et~al., 2012, \mn@doi [\aj] {10.1088/0004-6256/144/5/144},
  \href {https://ui.adsabs.harvard.edu/abs/2012AJ....144..144B} {144, 144}

\bibitem[\protect\citeauthoryear{{Booth} \& {Schaye}}{{Booth} \&
  {Schaye}}{2010}]{booth10}
{Booth} C.~M.,  {Schaye} J.,  2010, \mn@doi [\mnras]
  {10.1111/j.1745-3933.2010.00832.x}, \href
  {http://adsabs.harvard.edu/abs/2010MNRAS.405L...1B} {405, L1}

\bibitem[\protect\citeauthoryear{{Cisternas} et~al.,}{{Cisternas}
  et~al.}{2011}]{cisternas11}
{Cisternas} M.,  et~al., 2011, \mn@doi [\apjl] {10.1088/2041-8205/741/1/L11},
  \href {http://adsabs.harvard.edu/abs/2011ApJ...741L..11C} {741, L11}

\bibitem[\protect\citeauthoryear{{Dav{\'e}}, {Angl{\'e}s-Alc{\'a}zar},
  {Narayanan}, {Li}, {Rafieferantsoa}  \& {Appleby}}{{Dav{\'e}}
  et~al.}{2019a}]{Dave2019}
{Dav{\'e}} R.,  {Angl{\'e}s-Alc{\'a}zar} D.,  {Narayanan} D.,  {Li} Q.,
  {Rafieferantsoa} M.~H.,   {Appleby} S.,  2019a, \mn@doi [\mnras]
  {10.1093/mnras/stz937}, \href
  {https://ui.adsabs.harvard.edu/abs/2019MNRAS.486.2827D} {486, 2827}

\bibitem[\protect\citeauthoryear{{Dav{\'e}}, {Angl{\'e}s-Alc{\'a}zar},
  {Narayanan}, {Li}, {Rafieferantsoa}  \& {Appleby}}{{Dav{\'e}}
  et~al.}{2019b}]{dave19}
{Dav{\'e}} R.,  {Angl{\'e}s-Alc{\'a}zar} D.,  {Narayanan} D.,  {Li} Q.,
  {Rafieferantsoa} M.~H.,   {Appleby} S.,  2019b, \mn@doi [\mnras]
  {10.1093/mnras/stz937}, \href
  {https://ui.adsabs.harvard.edu/abs/2019MNRAS.486.2827D} {486, 2827}

\bibitem[\protect\citeauthoryear{{Davis}, {Graham}  \& {Cameron}}{{Davis}
  et~al.}{2019}]{davis19}
{Davis} B.~L.,  {Graham} A.~W.,   {Cameron} E.,  2019, \mn@doi [\apj]
  {10.3847/1538-4357/aaf3b8}, \href
  {https://ui.adsabs.harvard.edu/abs/2019ApJ...873...85D} {873, 85}

\bibitem[\protect\citeauthoryear{Ding et~al.,}{Ding et~al.}{2020}]{ding20}
Ding X.,  et~al., 2020, \mn@doi [The Astrophysical Journal]
  {10.3847/1538-4357/ab5b90}, 888, 37

\bibitem[\protect\citeauthoryear{{Du}, {Ho}, {Debattista}, {Pillepich},
  {Nelson}, {Hernquist}  \& {Weinberger}}{{Du} et~al.}{2021}]{du21}
{Du} M.,  {Ho} L.~C.,  {Debattista} V.~P.,  {Pillepich} A.,  {Nelson} D.,
  {Hernquist} L.,   {Weinberger} R.,  2021, \mn@doi [\apj]
  {10.3847/1538-4357/ac0e98}, \href
  {https://ui.adsabs.harvard.edu/abs/2021ApJ...919..135D} {919, 135}

\bibitem[\protect\citeauthoryear{Dubois \& Teyssier}{Dubois \&
  Teyssier}{2008}]{dubois08}
Dubois Y.,  Teyssier R.,  2008, \mn@doi [\aap] {10.1051/0004-6361:20078326},
  477, 79

\bibitem[\protect\citeauthoryear{Dubois, Devriendt, Slyz  \& Teyssier}{Dubois
  et~al.}{2012}]{dubois12}
Dubois Y.,  Devriendt J.,  Slyz A.,   Teyssier R.,  2012, \mn@doi [\mnras]
  {10.1111/j.1365-2966.2011.20236.x}, 420, 2662

\bibitem[\protect\citeauthoryear{{Dubois} et~al.,}{{Dubois}
  et~al.}{2014}]{dubois14}
{Dubois} Y.,  et~al., 2014, \mn@doi [\mnras] {10.1093/mnras/stu1227}, \href
  {https://ui.adsabs.harvard.edu/abs/2014MNRAS.444.1453D} {444, 1453}

\bibitem[\protect\citeauthoryear{{Dubois}, {Peirani}, {Pichon}, {Devriendt},
  {Gavazzi}, {Welker}  \& {Volonteri}}{{Dubois} et~al.}{2016}]{dubois16}
{Dubois} Y.,  {Peirani} S.,  {Pichon} C.,  {Devriendt} J.,  {Gavazzi} R.,
  {Welker} C.,   {Volonteri} M.,  2016, \mn@doi [\mnras]
  {10.1093/mnras/stw2265}, \href
  {https://ui.adsabs.harvard.edu/abs/2016MNRAS.463.3948D} {463, 3948}

\bibitem[\protect\citeauthoryear{{Gabor} \& {Dav{\'e}}}{{Gabor} \&
  {Dav{\'e}}}{2015}]{gabor15}
{Gabor} J.~M.,  {Dav{\'e}} R.,  2015, \mn@doi [\mnras] {10.1093/mnras/stu2399},
  \href {http://adsabs.harvard.edu/abs/2015MNRAS.447..374G} {447, 374}

\bibitem[\protect\citeauthoryear{{Gargiulo}, {Cora}, {Vega-Mart{\'\i}nez},
  {Gonzalez}, {Zoccali}, {Gonz{\'a}lez}, {Ruiz}  \& {Padilla}}{{Gargiulo}
  et~al.}{2017}]{gargiulo17}
{Gargiulo} I.~D.,  {Cora} S.~A.,  {Vega-Mart{\'\i}nez} C.~A.,  {Gonzalez}
  O.~A.,  {Zoccali} M.,  {Gonz{\'a}lez} R.,  {Ruiz} A.~N.,   {Padilla} N.~D.,
  2017, \mn@doi [\mnras] {10.1093/mnras/stx2188}, \href
  {https://ui.adsabs.harvard.edu/abs/2017MNRAS.472.4133G} {472, 4133}

\bibitem[\protect\citeauthoryear{{Granato}, {De Zotti}, {Silva}, {Bressan}  \&
  {Danese}}{{Granato} et~al.}{2004}]{granato04}
{Granato} G.~L.,  {De Zotti} G.,  {Silva} L.,  {Bressan} A.,   {Danese} L.,
  2004, \mn@doi [\apj] {10.1086/379875}, \href
  {https://ui.adsabs.harvard.edu/abs/2004ApJ...600..580G} {600, 580}

\bibitem[\protect\citeauthoryear{{Greene} et~al.,}{{Greene}
  et~al.}{2010}]{greene10b}
{Greene} J.~E.,  et~al., 2010, \mn@doi [\apj] {10.1088/0004-637X/721/1/26},
  \href {http://adsabs.harvard.edu/abs/2010ApJ...721...26G} {721, 26}

\bibitem[\protect\citeauthoryear{Greene, Strader  \& Ho}{Greene
  et~al.}{2020}]{greene20}
Greene J.~E.,  Strader J.,   Ho L.~C.,  2020, \mn@doi [Annual Review of
  Astronomy and Astrophysics] {10.1146/annurev-astro-032620-021835}, 58, 257

\bibitem[\protect\citeauthoryear{{G{\"u}ltekin} et~al.,}{{G{\"u}ltekin}
  et~al.}{2009}]{gultekin09}
{G{\"u}ltekin} K.,  et~al., 2009, \mn@doi [\apj] {10.1088/0004-637X/698/1/198},
  \href {http://adsabs.harvard.edu/abs/2009ApJ...698..198G} {698, 198}

\bibitem[\protect\citeauthoryear{{Habouzit} et~al.,}{{Habouzit}
  et~al.}{2021}]{Habouzit2021}
{Habouzit} M.,  et~al., 2021, \mn@doi [\mnras] {10.1093/mnras/stab496}, \href
  {https://ui.adsabs.harvard.edu/abs/2021MNRAS.503.1940H} {503, 1940}

\bibitem[\protect\citeauthoryear{{Habouzit} et~al.,}{{Habouzit}
  et~al.}{2022}]{Habouzit2022}
{Habouzit} M.,  et~al., 2022, \mn@doi [\mnras] {10.1093/mnras/stab3147}, \href
  {https://ui.adsabs.harvard.edu/abs/2022MNRAS.509.3015H} {509, 3015}

\bibitem[\protect\citeauthoryear{{H{\"a}ring} \& {Rix}}{{H{\"a}ring} \&
  {Rix}}{2004}]{haringrix04}
{H{\"a}ring} N.,  {Rix} H.-W.,  2004, \mn@doi [\apjl] {10.1086/383567}, \href
  {http://adsabs.harvard.edu/abs/2004ApJ...604L..89H} {604, L89}

\bibitem[\protect\citeauthoryear{{Heckman} \& {Best}}{{Heckman} \&
  {Best}}{2014}]{heckmanbest14}
{Heckman} T.~M.,  {Best} P.~N.,  2014, \mn@doi [\araa]
  {10.1146/annurev-astro-081913-035722}, \href
  {http://adsabs.harvard.edu/abs/2014ARA%26A..52..589H} {52, 589}

\bibitem[\protect\citeauthoryear{{Hopkins}, {Cox}, {Kere{\v s}}  \&
  {Hernquist}}{{Hopkins} et~al.}{2008}]{hopkins08b}
{Hopkins} P.~F.,  {Cox} T.~J.,  {Kere{\v s}} D.,   {Hernquist} L.,  2008,
  \mn@doi [\apjs] {10.1086/524363}, \href
  {http://adsabs.harvard.edu/abs/2008ApJS..175..390H} {175, 390}

\bibitem[\protect\citeauthoryear{{Hopkins}, {Quataert}  \& {Murray}}{{Hopkins}
  et~al.}{2012}]{hopkins12}
{Hopkins} P.~F.,  {Quataert} E.,   {Murray} N.,  2012, \mn@doi [\mnras]
  {10.1111/j.1365-2966.2012.20593.x}, \href
  {http://adsabs.harvard.edu/abs/2012MNRAS.tmp.2655H} {p.~2655}

\bibitem[\protect\citeauthoryear{{Hu}}{{Hu}}{2008}]{hu08}
{Hu} J.,  2008, \mn@doi [\mnras] {10.1111/j.1365-2966.2008.13195.x}, \href
  {http://adsabs.harvard.edu/abs/2008MNRAS.386.2242H} {386, 2242}

\bibitem[\protect\citeauthoryear{{Jahnke} \& {Macci{\`o}}}{{Jahnke} \&
  {Macci{\`o}}}{2011a}]{jahnke11}
{Jahnke} K.,  {Macci{\`o}} A.~V.,  2011a, \mn@doi [\apj]
  {10.1088/0004-637X/734/2/92}, \href
  {http://adsabs.harvard.edu/abs/2011ApJ...734...92J} {734, 92}

\bibitem[\protect\citeauthoryear{{Jahnke} \& {Macci{\`o}}}{{Jahnke} \&
  {Macci{\`o}}}{2011b}]{Knud2011}
{Jahnke} K.,  {Macci{\`o}} A.~V.,  2011b, \mn@doi [\apj]
  {10.1088/0004-637X/734/2/92}, \href
  {https://ui.adsabs.harvard.edu/abs/2011ApJ...734...92J} {734, 92}

\bibitem[\protect\citeauthoryear{{Jiang}, {Greene}, {Ho}, {Xiao}  \&
  {Barth}}{{Jiang} et~al.}{2011}]{jiang11b}
{Jiang} Y.-F.,  {Greene} J.~E.,  {Ho} L.~C.,  {Xiao} T.,   {Barth} A.~J.,
  2011, \mn@doi [\apj] {10.1088/0004-637X/742/2/68}, \href
  {http://adsabs.harvard.edu/abs/2011ApJ...742...68J} {742, 68}

\bibitem[\protect\citeauthoryear{{Kannappan} \& {Gawiser}}{{Kannappan} \&
  {Gawiser}}{2007}]{kannappan07}
{Kannappan} S.~J.,  {Gawiser} E.,  2007, \mn@doi [\apjl] {10.1086/512974},
  \href {https://ui.adsabs.harvard.edu/abs/2007ApJ...657L...5K} {657, L5}

\bibitem[\protect\citeauthoryear{{Kauffmann} et~al.,}{{Kauffmann}
  et~al.}{2003}]{kauffmann03}
{Kauffmann} G.,  et~al., 2003, \mn@doi [\mnras]
  {10.1046/j.1365-8711.2003.06291.x}, \href
  {http://adsabs.harvard.edu/abs/2003MNRAS.341...33K} {341, 33}

\bibitem[\protect\citeauthoryear{{Kaviraj} et~al.,}{{Kaviraj}
  et~al.}{2017}]{Kaviraj2017}
{Kaviraj} S.,  et~al., 2017, \mn@doi [\mnras] {10.1093/mnras/stx126}, \href
  {https://ui.adsabs.harvard.edu/abs/2017MNRAS.467.4739K} {467, 4739}

\bibitem[\protect\citeauthoryear{{Kelly}}{{Kelly}}{2007}]{kelly07}
{Kelly} B.~C.,  2007, \mn@doi [\apj] {10.1086/519947}, \href
  {http://adsabs.harvard.edu/abs/2007ApJ...665.1489K} {665, 1489}

\bibitem[\protect\citeauthoryear{{Kennicutt}}{{Kennicutt}}{1998}]{kennicutt98}
{Kennicutt} Jr. R.~C.,  1998, \mn@doi [\apj] {10.1086/305588}, \href
  {http://adsabs.harvard.edu/abs/1998ApJ...498..541K} {498, 541}

\bibitem[\protect\citeauthoryear{Kimm, Cen, Devriendt, Dubois  \& Slyz}{Kimm
  et~al.}{2015}]{Kimm15}
Kimm T.,  Cen R.,  Devriendt J.,  Dubois Y.,   Slyz A.,  2015, \mn@doi [\mnras]
  {10.1093/mnras/stv1211}, 451, 2900

\bibitem[\protect\citeauthoryear{{Kocevski} et~al.,}{{Kocevski}
  et~al.}{2012}]{kocevski12}
{Kocevski} D.~D.,  et~al., 2012, \mn@doi [\apj] {10.1088/0004-637X/744/2/148},
  \href {http://adsabs.harvard.edu/abs/2012ApJ...744..148K} {744, 148}

\bibitem[\protect\citeauthoryear{{Komatsu} et~al.,}{{Komatsu}
  et~al.}{2011}]{komatsu11}
{Komatsu} E.,  et~al., 2011, \mn@doi [\apjs] {10.1088/0067-0049/192/2/18},
  \href {http://adsabs.harvard.edu/abs/2011ApJS..192...18K} {192, 18}

\bibitem[\protect\citeauthoryear{{Kormendy}, {Drory}, {Bender}  \&
  {Cornell}}{{Kormendy} et~al.}{2010}]{kormendy10}
{Kormendy} J.,  {Drory} N.,  {Bender} R.,   {Cornell} M.~E.,  2010, \mn@doi
  [\apj] {10.1088/0004-637X/723/1/54}, \href
  {http://adsabs.harvard.edu/abs/2010ApJ...723...54K} {723, 54}

\bibitem[\protect\citeauthoryear{{Kormendy}, {Bender}  \& {Cornell}}{{Kormendy}
  et~al.}{2011}]{kormendy11a}
{Kormendy} J.,  {Bender} R.,   {Cornell} M.~E.,  2011, \mn@doi [\nat]
  {10.1038/nature09694}, \href
  {http://adsabs.harvard.edu/abs/2011Natur.469..374K} {469, 374}

\bibitem[\protect\citeauthoryear{{Koss}, {Mushotzky}, {Veilleux}, {Winter},
  {Baumgartner}, {Tueller}, {Gehrels}  \& {Valencic}}{{Koss}
  et~al.}{2011}]{koss11}
{Koss} M.,  {Mushotzky} R.,  {Veilleux} S.,  {Winter} L.~M.,  {Baumgartner} W.,
   {Tueller} J.,  {Gehrels} N.,   {Valencic} L.,  2011, \mn@doi [\apj]
  {10.1088/0004-637X/739/2/57}, \href
  {http://adsabs.harvard.edu/abs/2011ApJ...739...57K} {739, 57}

\bibitem[\protect\citeauthoryear{{Madau}, {Pozzetti}  \& {Dickinson}}{{Madau}
  et~al.}{1998}]{madau98}
{Madau} P.,  {Pozzetti} L.,   {Dickinson} M.,  1998, \mn@doi [\apj]
  {10.1086/305523}, \href {http://adsabs.harvard.edu/abs/1998ApJ...498..106M}
  {498, 106}

\bibitem[\protect\citeauthoryear{{Magorrian} et~al.,}{{Magorrian}
  et~al.}{1998}]{magorrian98}
{Magorrian} J.,  et~al., 1998, \mn@doi [\aj] {10.1086/300353}, \href
  {http://adsabs.harvard.edu/cgi-bin/nph-bib_query?bibcode=1998AJ....115.2285M&db_key=AST}
  {115, 2285}

\bibitem[\protect\citeauthoryear{{Marconi} \& {Hunt}}{{Marconi} \&
  {Hunt}}{2003}]{marconi03}
{Marconi} A.,  {Hunt} L.~K.,  2003, \mn@doi [\apjl] {10.1086/375804}, \href
  {http://adsabs.harvard.edu/abs/2003ApJ...589L..21M} {589, L21}

\bibitem[\protect\citeauthoryear{{Martig}, {Bournaud}, {Croton}, {Dekel}  \&
  {Teyssier}}{{Martig} et~al.}{2012}]{martig12}
{Martig} M.,  {Bournaud} F.,  {Croton} D.~J.,  {Dekel} A.,   {Teyssier} R.,
  2012, \mn@doi [\apj] {10.1088/0004-637X/756/1/26}, \href
  {http://adsabs.harvard.edu/abs/2012ApJ...756...26M} {756, 26}

\bibitem[\protect\citeauthoryear{{Martin} et~al.,}{{Martin}
  et~al.}{2018a}]{martin18}
{Martin} G.,  et~al., 2018a, \mn@doi [\mnras] {10.1093/mnras/sty324}, \href
  {https://ui.adsabs.harvard.edu/\#abs/2018MNRAS.476.2801M} {476, 2801}

\bibitem[\protect\citeauthoryear{{Martin}, {Kaviraj}, {Devriendt}, {Dubois}  \&
  {Pichon}}{{Martin} et~al.}{2018b}]{martin18b}
{Martin} G.,  {Kaviraj} S.,  {Devriendt} J.~E.~G.,  {Dubois} Y.,   {Pichon} C.,
   2018b, \mn@doi [\mnras] {10.1093/mnras/sty1936}, \href
  {https://ui.adsabs.harvard.edu/abs/2018MNRAS.480.2266M} {480, 2266}

\bibitem[\protect\citeauthoryear{{McAlpine}, {Harrison}, {Rosario},
  {Alexander}, {Ellison}, {Johansson}  \& {Patton}}{{McAlpine}
  et~al.}{2020}]{mcalpine20}
{McAlpine} S.,  {Harrison} C.~M.,  {Rosario} D.~J.,  {Alexander} D.~M.,
  {Ellison} S.~L.,  {Johansson} P.~H.,   {Patton} D.~R.,  2020, \mn@doi
  [\mnras] {10.1093/mnras/staa1123}, \href
  {https://ui.adsabs.harvard.edu/abs/2020MNRAS.494.5713M} {494, 5713}

\bibitem[\protect\citeauthoryear{{McConnell} \& {Ma}}{{McConnell} \&
  {Ma}}{2013}]{mcconnell13}
{McConnell} N.~J.,  {Ma} C.-P.,  2013, \mn@doi [\apj]
  {10.1088/0004-637X/764/2/184}, \href
  {https://ui.adsabs.harvard.edu/abs/2013ApJ...764..184M} {764, 184}

\bibitem[\protect\citeauthoryear{{Merritt} \& {Ferrarese}}{{Merritt} \&
  {Ferrarese}}{2001}]{merritt01}
{Merritt} D.,  {Ferrarese} L.,  2001, \mn@doi [\mnras]
  {10.1046/j.1365-8711.2001.04165.x}, \href
  {http://adsabs.harvard.edu/abs/2001MNRAS.320L..30M} {320, L30}

\bibitem[\protect\citeauthoryear{Ostriker}{Ostriker}{1999}]{ostriker99}
Ostriker E.~C.,  1999, \mn@doi [\apj] {10.1086/306858}, 513, 252

\bibitem[\protect\citeauthoryear{{Parry}, {Eke}  \& {Frenk}}{{Parry}
  et~al.}{2009}]{parry09}
{Parry} O.~H.,  {Eke} V.~R.,   {Frenk} C.~S.,  2009, \mn@doi [\mnras]
  {10.1111/j.1365-2966.2009.14921.x}, \href
  {http://adsabs.harvard.edu/abs/2009MNRAS.396.1972P} {396, 1972}

\bibitem[\protect\citeauthoryear{{Peng}}{{Peng}}{2007}]{peng07}
{Peng} C.~Y.,  2007, \apj, 671, 1098

\bibitem[\protect\citeauthoryear{{Povi{\'c}} et~al.,}{{Povi{\'c}}
  et~al.}{2012}]{povic12}
{Povi{\'c}} M.,  et~al., 2012, \mn@doi [\aap] {10.1051/0004-6361/201117314},
  \href {https://ui.adsabs.harvard.edu/abs/2012A&A...541A.118P} {541, A118}

\bibitem[\protect\citeauthoryear{{Rakshit} \& {Woo}}{{Rakshit} \&
  {Woo}}{2018}]{RW18}
{Rakshit} S.,  {Woo} J.-H.,  2018, \mn@doi [\apj] {10.3847/1538-4357/aad9f8},
  \href {https://ui.adsabs.harvard.edu/abs/2018ApJ...865....5R} {865, 5}

\bibitem[\protect\citeauthoryear{{Reines} \& {Volonteri}}{{Reines} \&
  {Volonteri}}{2015}]{reines15}
{Reines} A.~E.,  {Volonteri} M.,  2015, \mn@doi [\apj]
  {10.1088/0004-637X/813/2/82}, \href
  {https://ui.adsabs.harvard.edu/abs/2015ApJ...813...82R} {813, 82}

\bibitem[\protect\citeauthoryear{{Saglia} et~al.,}{{Saglia}
  et~al.}{2016}]{saglia16}
{Saglia} R.~P.,  et~al., 2016, \mn@doi [\apj] {10.3847/0004-637X/818/1/47},
  \href {https://ui.adsabs.harvard.edu/abs/2016ApJ...818...47S} {818, 47}

\bibitem[\protect\citeauthoryear{{Sahu}, {Graham}  \& {Davis}}{{Sahu}
  et~al.}{2019}]{sahu19}
{Sahu} N.,  {Graham} A.~W.,   {Davis} B.~L.,  2019, \mn@doi [\apj]
  {10.3847/1538-4357/ab0f32}, \href
  {https://ui.adsabs.harvard.edu/abs/2019ApJ...876..155S} {876, 155}

\bibitem[\protect\citeauthoryear{{Salpeter}}{{Salpeter}}{1955}]{salpeter55}
{Salpeter} E.~E.,  1955, \mn@doi [\apj] {10.1086/145971}, \href
  {http://adsabs.harvard.edu/abs/1955ApJ...121..161S} {121, 161}

\bibitem[\protect\citeauthoryear{{Salucci}, {Ratnam}, {Monaco}  \&
  {Danese}}{{Salucci} et~al.}{2000}]{salucci00}
{Salucci} P.,  {Ratnam} C.,  {Monaco} P.,   {Danese} L.,  2000, \mn@doi
  [\mnras] {10.1046/j.1365-8711.2000.03622.x}, \href
  {https://ui.adsabs.harvard.edu/abs/2000MNRAS.317..488S} {317, 488}

\bibitem[\protect\citeauthoryear{{Schawinski} et~al.,}{{Schawinski}
  et~al.}{2010}]{schawinski10a}
{Schawinski} K.,  et~al., 2010, \mn@doi [\apj] {10.1088/0004-637X/711/1/284},
  \href {http://adsabs.harvard.edu/abs/2010ApJ...711..284S} {711, 284}

\bibitem[\protect\citeauthoryear{Sijacki, Vogelsberger, Genel, Springel,
  Torrey, Snyder, Nelson  \& Hernquist}{Sijacki et~al.}{2015}]{sijacki15}
Sijacki D.,  Vogelsberger M.,  Genel S.,  Springel V.,  Torrey P.,  Snyder
  G.~F.,  Nelson D.,   Hernquist L.,  2015, \mn@doi [Monthly Notices of the
  Royal Astronomical Society] {10.1093/mnras/stv1340}, 452, 575

\bibitem[\protect\citeauthoryear{{Silk} \& {Rees}}{{Silk} \&
  {Rees}}{1998}]{silk98}
{Silk} J.,  {Rees} M.~J.,  1998, \aap, \href
  {http://adsabs.harvard.edu/abs/1998A%26A...331L...1S} {331, L1}

\bibitem[\protect\citeauthoryear{{Simmons}, {Van Duyne}, {Urry}, {Treister},
  {Koekemoer}, {Grogin}  \& {The GOODS Team}}{{Simmons}
  et~al.}{2011}]{simmons11}
{Simmons} B.~D.,  {Van Duyne} J.,  {Urry} C.~M.,  {Treister} E.,  {Koekemoer}
  A.~M.,  {Grogin} N.~A.,   {The GOODS Team} 2011, \mn@doi [\apj]
  {10.1088/0004-637X/734/2/121}, \href
  {http://adsabs.harvard.edu/abs/2011ApJ...734..121S} {734, 121}

\bibitem[\protect\citeauthoryear{{Simmons} et~al.,}{{Simmons}
  et~al.}{2013}]{simmons13}
{Simmons} B.~D.,  et~al., 2013, \mn@doi [\mnras] {10.1093/mnras/sts491}, \href
  {http://adsabs.harvard.edu/abs/2013MNRAS.429.2199S} {429, 2199}

\bibitem[\protect\citeauthoryear{{Simmons}, {Smethurst}  \&
  {Lintott}}{{Simmons} et~al.}{2017}]{ssl17}
{Simmons} B.~D.,  {Smethurst} R.~J.,   {Lintott} C.,  2017, \mn@doi [\mnras]
  {10.1093/mnras/stx1340}, \href
  {https://ui.adsabs.harvard.edu/\#abs/2017MNRAS.470.1559S} {470, 1559}

\bibitem[\protect\citeauthoryear{{Smethurst} et~al.,}{{Smethurst}
  et~al.}{2016}]{smethurst16}
{Smethurst} R.~J.,  et~al., 2016, \mn@doi [\mnras] {10.1093/mnras/stw2204},
  \href {http://adsabs.harvard.edu/abs/2016MNRAS.463.2986S} {463, 2986}

\bibitem[\protect\citeauthoryear{{Smethurst}, {Simmons}, {Lintott}  \&
  {Shanahan}}{{Smethurst} et~al.}{2019}]{smethurst19}
{Smethurst} R.~J.,  {Simmons} B.~D.,  {Lintott} C.~J.,   {Shanahan} J.,  2019,
  \mn@doi [\mnras] {10.1093/mnras/stz2443}, \href
  {https://ui.adsabs.harvard.edu/abs/2019MNRAS.489.4016S} {489, 4016}

\bibitem[\protect\citeauthoryear{{Smethurst} et~al.,}{{Smethurst}
  et~al.}{2021}]{smethurst21}
{Smethurst} R.~J.,  et~al., 2021, \mn@doi [\mnras] {10.1093/mnras/stab2340},
  \href {https://ui.adsabs.harvard.edu/abs/2021MNRAS.507.3985S} {507, 3985}

\bibitem[\protect\citeauthoryear{Steinborn, Hirschmann, Dolag, Shankar, Juneau,
  Krumpe, Remus  \& Teklu}{Steinborn et~al.}{2018}]{steinborn18}
Steinborn L.~K.,  Hirschmann M.,  Dolag K.,  Shankar F.,  Juneau S.,  Krumpe
  M.,  Remus R.-S.,   Teklu A.~F.,  2018, \mn@doi [Monthly Notices of the Royal
  Astronomical Society] {10.1093/mnras/sty2288}, 481, 341

\bibitem[\protect\citeauthoryear{{Teyssier}}{{Teyssier}}{2002}]{teyssier02}
{Teyssier} R.,  2002, \mn@doi [\aap] {10.1051/0004-6361:20011817}, \href
  {https://ui.adsabs.harvard.edu/abs/2002A&A...385..337T} {385, 337}

\bibitem[\protect\citeauthoryear{{Tonini}, {Mutch}, {Croton}  \&
  {Wyithe}}{{Tonini} et~al.}{2016}]{tonini16}
{Tonini} C.,  {Mutch} S.~J.,  {Croton} D.~J.,   {Wyithe} J.~S.~B.,  2016,
  \mn@doi [\mnras] {10.1093/mnras/stw956}, \href
  {http://ukads.nottingham.ac.uk/abs/2016MNRAS.459.4109T} {459, 4109}

\bibitem[\protect\citeauthoryear{Torrey et~al.,}{Torrey
  et~al.}{2020}]{torrey20}
Torrey P.,  et~al., 2020, \mn@doi [Monthly Notices of the Royal Astronomical
  Society] {10.1093/mnras/staa2222}, 497, 5292

\bibitem[\protect\citeauthoryear{Tweed, Devriendt, Blaizot, Colombi  \&
  Slyz}{Tweed et~al.}{2009}]{tweed2009}
Tweed D.,  Devriendt J.,  Blaizot J.,  Colombi S.,   Slyz A.,  2009, \mn@doi
  [\aap] {10.1051/0004-6361/200911787}, 506, 647

\bibitem[\protect\citeauthoryear{{Villforth} et~al.,}{{Villforth}
  et~al.}{2014}]{villforth14}
{Villforth} C.,  et~al., 2014, \mn@doi [\mnras] {10.1093/mnras/stu173}, \href
  {https://ui.adsabs.harvard.edu/abs/2014MNRAS.439.3342V} {439, 3342}

\bibitem[\protect\citeauthoryear{Volonteri, Dubois, Pichon  \&
  Devriendt}{Volonteri et~al.}{2016}]{volonteri16}
Volonteri M.,  Dubois Y.,  Pichon C.,   Devriendt J.,  2016, \mn@doi [\mnras]
  {10.1093/mnras/stw1123}, 460, 2979

\bibitem[\protect\citeauthoryear{{Walker}, {Mihos}  \& {Hernquist}}{{Walker}
  et~al.}{1996}]{walker96}
{Walker} I.~R.,  {Mihos} J.~C.,   {Hernquist} L.,  1996, \mn@doi [\apj]
  {10.1086/176956}, \href {http://adsabs.harvard.edu/abs/1996ApJ...460..121W}
  {460, 121}

\bibitem[\protect\citeauthoryear{{Weinberger} et~al.,}{{Weinberger}
  et~al.}{2017}]{Weinberger2017}
{Weinberger} R.,  et~al., 2017, \mn@doi [\mnras] {10.1093/mnras/stw2944}, \href
  {https://ui.adsabs.harvard.edu/abs/2017MNRAS.465.3291W} {465, 3291}

\bibitem[\protect\citeauthoryear{{York} et~al.,}{{York} et~al.}{2000}]{york00}
{York} D.~G.,  et~al., 2000, \mn@doi [\aj] {10.1086/301513}, \href
  {http://adsabs.harvard.edu/abs/2000AJ....120.1579Y} {120, 1579}

\bibitem[\protect\citeauthoryear{{Zhao}, {Ho}, {Shangguan}, {Kim}, {Zhao}  \&
  {Gao}}{{Zhao} et~al.}{2021}]{zhao21}
{Zhao} Y.,  {Ho} L.~C.,  {Shangguan} J.,  {Kim} M.,  {Zhao} D.,   {Gao} H.,
  2021, \mn@doi [\apj] {10.3847/1538-4357/abe8d4}, \href
  {https://ui.adsabs.harvard.edu/abs/2021ApJ...911...94Z} {911, 94}

\bibitem[\protect\citeauthoryear{{Zhong}, {Inoue}, {Yamanaka}  \&
  {Yamada}}{{Zhong} et~al.}{2022}]{zhong22}
{Zhong} Y.,  {Inoue} A.~K.,  {Yamanaka} S.,   {Yamada} T.,  2022, \mn@doi
  [\apj] {10.3847/1538-4357/ac3edb}, \href
  {https://ui.adsabs.harvard.edu/abs/2022ApJ...925..157Z} {925, 157}

\bibitem[\protect\citeauthoryear{{van den Bosch}}{{van den
  Bosch}}{2016}]{vandenbosch16}
{van den Bosch} R. C.~E.,  2016, \mn@doi [\apj] {10.3847/0004-637X/831/2/134},
  \href {https://ui.adsabs.harvard.edu/abs/2016ApJ...831..134V} {831, 134}

\makeatother
\end{thebibliography}








\bsp	
\label{lastpage}
\end{document}